\documentclass{emulateapj}
\shorttitle{Electron dissipation range of solar wind turbulence }
\shortauthors{Sahraoui et al.}

\begin{document}


\title{SCALING OF THE ELECTRON DISSIPATION RANGE OF SOLAR WIND TURBULENCE}


\author{F. Sahraoui\altaffilmark{1}, S. Y. Huang\altaffilmark{2}, J. De Patoul\altaffilmark{1}, G. Belmont\altaffilmark{1}, M. L. Goldstein\altaffilmark{3}, A. R\'etino\altaffilmark{1}, P. Robert\altaffilmark{1}, N. Cornilleau-Wehrlin\altaffilmark{1}}
\affil{\altaffilmark{1}Laboratoire de Physique des Plasmas, CNRS-Ecole Polytechnique-UPMC, Route de Saclay, 92120 Palaiseau, France}
\affil{\altaffilmark{2}School of Electronics and Information, Wuhan University, Wuhan, China}
\affil{\altaffilmark{3}NASA Goddard Space Flight Center, Code 673, Greenbelt 20771, Maryland, USA}
\email{fouad.sahraoui@lpp.polytechnique.fr}

%\altaffiltext{1}{Visiting Astronomer, Cerro Tololo Inter-American Observatory.
%CTIO is operated by AURA, Inc.\ under contract to the National Science
%\altaffiltext{2}{Society of Fellows, Harvard University.}

\begin{abstract}
Electron scale solar wind turbulence has attracted great interest in recent years. Clear evidences have been given from the Cluster data that turbulence is not fully dissipated near the proton scale but continues cascading down to the electron scales. However, the scaling of the energy spectra as well as the nature of the plasma modes involved at those small scales are still not fully determined. Here we survey 10 years of the Cluster search-coil magnetometer (SCM) waveforms measured in the solar wind and perform a statistical study of the magnetic energy spectra in the frequency range [$1, 180$]Hz. We show that a large fraction of the spectra exhibit clear breakpoints near the electron gyroscale $\rho_e$, followed by steeper power-law like spectra. We show that the scaling below the electron breakpoint cannot be determined unambiguously due to instrumental limitations that will be discussed in detail. We compare our results to those reported in other studies and discuss their implication on the physical mechanisms and the theoretical modeling of energy dissipation in the SW.
\end{abstract}

\keywords{turbulence, solar wind, heating, whistler, KAW}

%--------------------------------------------------------------------------------------------
% New section 
%--------------------------------------------------------------------------------------------
\section{Introduction}
The Solar Wind (SW) is certainly the astrophysical plasma in which significant progress in understanding plasma turbulence has been most achieved. The main reason for that is the availability of high quality data from several space missions, from the earlier spacecraft Voyager 1 and 2 to more recent ones such as Cluster, Wind or Stereo. 
While earlier spacecraft data have allowed for addressing turbulence at scales larger than the ion scale (typically $\rho_i \sim 100$ km, which corresponds to an observed frequency in the spacecraft frame of $f_{sc} \sim 0.5$ Hz using the Taylor assumption), i.e. the so-called inertial range, the high time resolution Cluster data have recently opened a new chapter in turbulence studies that focuses on electron scales ($L\sim\rho_e\sim 1$ km)~\citep{sahraoui09,kiyani09,alexandrova09,sahraoui10a}. These observations have driven intensive research work on electron scale turbulence both theoretically and numerically~\citep{chang11,meyrand10,podesta10,camporeale11,howes11a, sahraoui12, meyrand12}. Determining the nature and properties (e.g., scaling, anisotropy) of the turbulence at small scales is indeed crucial to understanding the problems of energy dissipation and heating, particle acceleration, and magnetic reconnection in space and astrophysical plasmas \citep{schekochihin09}. Thanks to the high time resolution Search-Coil Magnetometer (SCM), which provides waveforms of the magnetic field fluctuations up to $180$Hz, it has become clear that SW turbulence does cascade below the ion scale $\rho_i$ down to the electron scale $\rho_e$ where dissipation becomes important and the spectra steepen significantly. However, the underlying physics is still very controversial, owing to the fact that the available observations are very few and the theoretical and numerical work has been extended to those small scales only very recently ~\citep{chang11,camporeale11,howes11a,sahraoui12, meyrand12}. 

From the observational point of view only a handful of observations exist on electron scales SW turbulence and all do not agree. The first point of controversy is the scaling of the magnetic energy spectra down to and below $\rho_e$. \cite{sahraoui09} have first reported a power-law cascade $f^{-2.8}$ down to $f_{\rho_e}$, where a clear spectral break is observed and followed by another power-law like shown to be close to ${f_{sc}}^{-4}$ ($f_{\rho_e}$ is the frequency in the spacecraft frame corresponding to the electron gyroscale when the Taylor frozen-in-flow assumption is used\footnote{It is commonly referred to as the Doppler-shifted electron gyroradius, although it is not a very appropriate definition considering that Doppler shift applies to frequencies and not to spatial scales. Here we introduce the terminology of the ``Taylor-shifted scale".}). It was emphasized in that reference that the fit below $f_{\rho_e}$ is subject to caution owing to the limited extension of the electron dissipation range, imposed by the sensitivity of the SCM (see section \ref{caveats} below). Similar power-law-like spectra have been reported in~\cite{sahraoui10a, kiyani09}. On the other hand \cite{alexandrova09}, using STAFF-Spectrum Analyzer (STAFF-SA) data, have reported exponential scaling $\sim exp(-\sqrt{k\rho_e})$ in the scale range $k\rho_e \sim [0.1,1]$. More recently \cite{alexandrova12} proposed a different model where a scaling ${k_\perp}^{-8/3}exp{(-k_\perp\rho_e})$ was suggested to fit better the data in the scale range $k\rho_e \sim [0.03,3]$. We will refer to the above three models  respectively by the {\it double-power-law}, the {\it exponential} and the {\it hybrid}. Besides these three models we introduce below a new one, the {\it asymptotic-double-power-law}. We show that the exponential model does not fit well the data as does the double-power-law model. We show however that, due to Cluster instrumental limitations that we shall discuss in detail, this latter may not be always possible to distinguish within the data from the two remaining models, the main difference between them being reduced to the presence or not of the spectral break near $\rho_e$.

From the theoretical point of view, very recent numerical simulations have tried to tackle the problem of cascade and dissipation at electrons scales. ~\cite{camporeale11} performed 2D full Particle-In-Cell (PIC) simulations of decaying electromagnetic fluctuations near and below the electron inertial length $d_e$. They showed that the magnetic energy spectra steepen from $k_\perp^{-2.6}$ to $k_\perp^{-5.8}$ with a clear spectral break at $k\rho_e\sim 1$. 3D PIC simulations of whistler-driven turbulence showed very similar results with less steep spectra, ${k_\perp}^{-4.3}$, at scales $k_\perp d_e\gtrsim1$ \citep{chang11}. GyroKinetic (GK) simulations of strong Kinetic Alfv\'en Wave (KAW) turbulence, which self-consistently contain kinetic damping of the low frequency plasma modes ($\omega<<\omega_{ci}$) but the cyclotron resonance, showed clear power-law cascade $k_\perp^{-2.8}$ down to $k_\perp\rho_e\sim 1$ ~\citep{howes11a}. The GK theory predicts magnetic energy spectra with a scaling $k_\perp^{-16/3}$ for the (entropy) cascade below $\rho_e$~\citep{schekochihin09}. On the other hand,  \cite{meyrand10} showed that incompressible Electron MHD (EMHD) turbulence predicts that the magnetic energy spectra should follow a power law $k^{-11/3}$ at scales smaller than $d_e$. That steepening was proposed to explain the observed spectra below the electron scale reported in the SW \citep{sahraoui09,sahraoui10a}. Although this fluid model is non-dissipative and does not consider any damping of the turbulence via kinetic effects, which are important in the dispersive and the dissipation ranges, it has nevertheless the merit of proving that fully nonlinear dynamics and dispersive effects in EMHD leads to a steepening of the energy spectra near $d_e$. Thus, all existing theoretical and numerical predictions support the double-power-law model to reproduce the scaling near and below the electron gyroscale or inertial length. In a recent paper by \cite{howes11c} it was pointed out that for $\beta_i<<1$ the kinetic damping of the KAW turbulence becomes strong, which results in a cut-off of the spectra at moderate wavenumber, typically $k_\perp\rho_i\sim10$. This is inconsistent with SW observations~\citep{sahraoui09, sahraoui10a, alexandrova09}. To overcome the failure of the model to reproduce SW observations it was suggested in that paper that nonlocal effects, due to large scale shear flows, play a role in sustaining the energy cascade down to smaller scales. When such nonlocal effects are considered it was indeed shown that the energy cascade in low $\beta_i$ can reach scales $ k_\perp\rho_e\sim 1$, where the spectra curve and show an exponential-like behaviour. For $\beta_i\gtrsim 1$ relevant to the SW both the local and the nonlocal models give the same power-law scaling of the magnetic energy spectra down to $k_\perp\rho_e \sim1$ ~\citep{howes11a,howes11c}, which may indicate that nonlocal effects do not play an important role in this case . We will return to this discussion below.  

Here we report a large statistical survey of SW turbulence using the Cluster SCM data sampled at $450$Hz. We focus particularly on the scaling of the magnetic energy spectra up to and above $f_{\rho_e}$,  the frequency corresponding to Taylor-shifting the electron gyroscale $\rho_e$.  In section \ref{caveats} we discuss several caveats that need to be handled carefully in order to avoid any misinterpretation of the observations. In section \ref{data} we explain the approach we used to select the Cluster data in the SW. We particularly target intervals when the Cluster fleet is located in the SW, excluding thus data from the electron and ion foreshock regions, and when the Signal-to-Noise-Ratio (SNR) of the magnetic fluctuations is high. In section \ref{results} we show the main results of the study and discuss them in light of existing theoretical and numerical predictions and earlier work related to the subject.

%-----------------------------------------------------------------------------------------------------------------------------------------------------------------
% %%%%%%%%%%%%%%%%%%%%          New section 		%%%%%%%%%%%%%%%%%%%%%%%%
%-----------------------------------------------------------------------------------------------------------------------------------------------------------------
\section{Experimental caveats}\label{caveats}

%---------------------------------------------------------------------
% Subsection I
%---------------------------------------------------------------------
\subsection{Effect of the limited sensitivity of the SCM}\label{noise}
The Cluster SCM data are so far the only available data allowing one to probe into electron scales of SW turbulence with relatively high SNR. The SNR (in dB) is defined as a function of the measured frequency onboard the spacecraft~\citep{sahraoui10c}
\begin{displaymath}
SNR(f_{sc})=10Log_{10}\Bigg[\frac{\delta B^2(f_{sc})}{{\delta B_{sens.}}^2(f_{sc})}\Bigg]
\end{displaymath}
where $\delta B$ and $\delta B_{sens.}$ are respectively the amplitude of the magnetic fluctuation and the level of the sensitivity floor of the Cluster SCM at the frequency $f_{sc}$.
Current space missions (e.g. Stereo, Wind, Themis) and the planned ones (e.g. MMS) either do not have SCM or do have but with limited sensitivity. The sensitivity of the SCM is essentially (but not exclusively) limited by the length of the magnetic sensor, which is generally constrained by the payload on the spacecraft. For instance the SCMs onboard Cluster, Themis and MMS have sensors with respective lengths of $27$cm, $15$cm and  $10$cm which yields respective sensitivity levels of  $0.43$, $0.70$, and $2.35pT/\sqrt{Hz}$@$10$Hz. Despite this relatively high sensitivity of the Cluster SCM compared to all other space missions, it does not allow us all the time to measure the SW magnetic fluctuations above $\sim 30$Hz due to their very low amplitude (see Figs. 7 and 8 in \cite{sahraoui10c}). Indeed, when the measured SNR is exceptionally high (typically SNR$\gtrsim30$) one should suspect the crossings of other regions  such as the ion or electron foreshock \citep{sahraoui09} or boundaries such as interplanetary shocks or CMEs (e.g. the event of 2004-01-22 studied in \cite{alexandrova09,alexandrova12}). We recall that Cluster, Themis and MMS are magnetospheric missions and have been designed to address essentially the physics of different regions of the Earth's magnetosphere and the magnetosheath where the magnetic fluctuations have generally higher amplitudes than in the SW.

It follows from this discussion that in order to address properly the problem of energy cascade and dissipation in the SW  at electron scales, typically at frequencies $f_{sc} \gtrsim30$Hz, it is important to select SW data when the fluctuations amplitude is very high compared to the sensitivity floor of the Cluster SCM. Because of this limited SNR in the SW ~\cite{alexandrova09} proposed to subtract the sensitivity level of the instrument from the measured power spectra. Without discussing in detail the foundation of a such procedure from the signal processing point of view (in particular since the noise of the instrument is not known accurately, and is generally estimated from measurements in the magnetospheric lobes), we only underline how affected would be the scaling of the resulting ``denoised" spectra. Indeed, when the SNR becomes very low at a given frequency $f_0$, i.e. $\delta B^2(f \gtrsim f_0)=(1+\epsilon(f)){\delta B_{sens}}^2(f\gtrsim f_0)$ with $\epsilon(f)<<1$, subtracting the sensitivity of the instrument from the actual spectrum yields a spectrum which falls to zero:  $\delta B^2-{\delta B_{sens}}^2\sim \epsilon{\delta B_{sens}}^2$. It is not surprising that on the Log-Log scale such a vanishing spectrum will be seen as exponentially falling-off for all frequencies $f\gtrsim f_0$. 

%-------------------------------------------------------------------------
% Subsection II
%-------------------------------------------------------------------------
\subsection{Effect of the limited frequency resolution of STAFF-Spectrum Analyzer}\label{staff-sa}
\begin{figure}
\includegraphics[height=6.5cm,width=7.5cm]{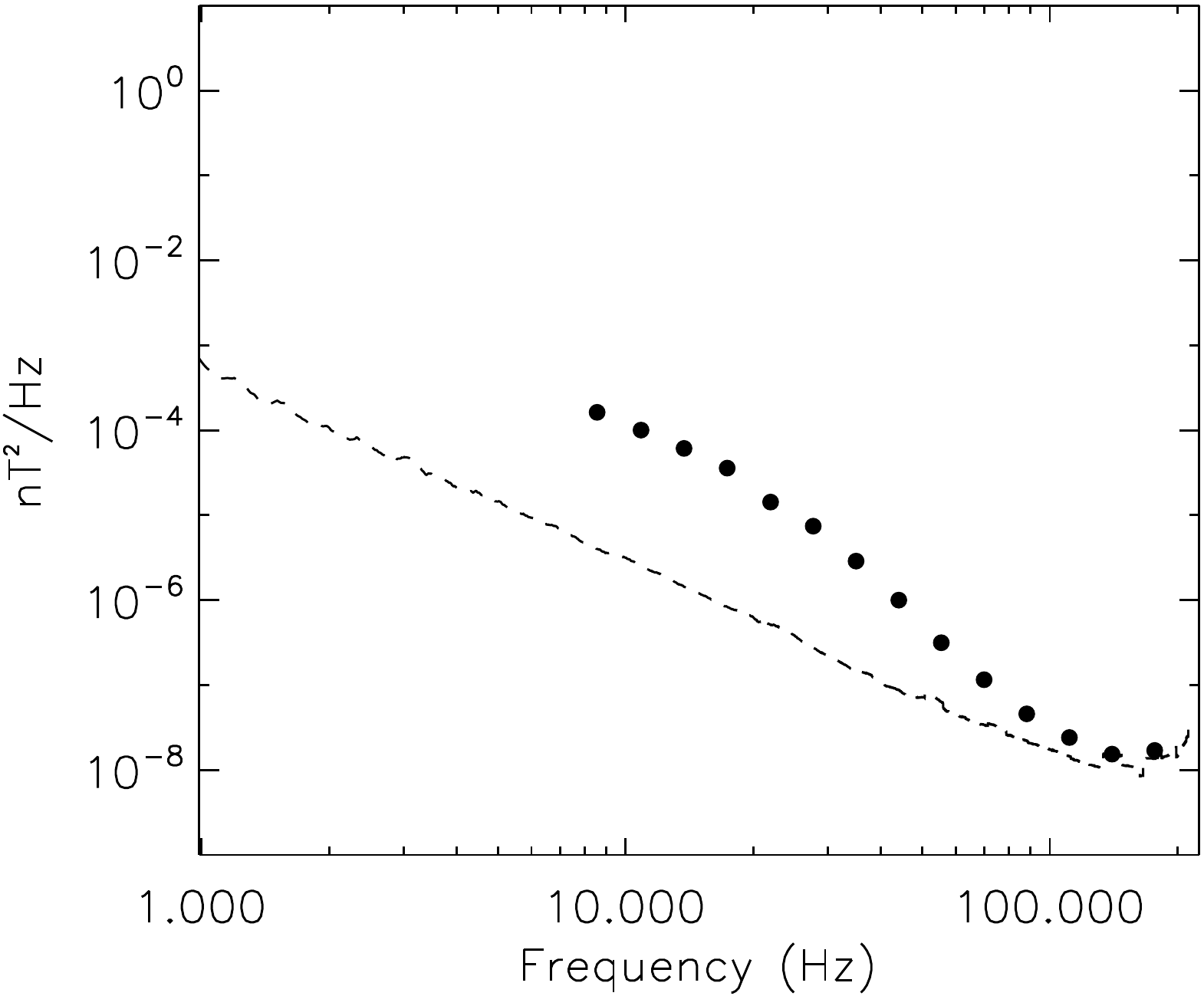}
\caption{The spectrum of Fig.\ref{spectra}(c) as it would have been measured by STAFF-SA with the available sampling frequencies. \label{sa}}
\end{figure}

The STAFF instrument onboard Cluster has a single tri-axial Search-Coil Magnetometer (SCM), but has two sub-experiments: STAFF-SC measures magnetic field fluctuations (i.e. waveforms) with two possible sampling time: 25 samples/sec  (Normal Mode, NM) and 450 samples/sec (Burst Mode, BM) with however a low-pass filter cut-off at $180$Hz, and STAFF-SA computes onboard the spacecraft the $5 \times 5$ spectral matrix of the three components of the magnetic field and the two components of the electric field fluctuations measured by EFW \citep{cornilleau03}. STAFF-SA transmits the computed correlation matrix on the central frequencies $f_0$ of $27$ channels logarithmically spaced in the frequency range [$8$Hz, $4$kHz] whose bandwidth is proportional to $f_0$, $\Delta f=13\%f_0 $. In BM the interval reduces to approximately [$70$Hz,$4$kHz]. It is important to emphasize here that due to the limited SNR in the SW, which generally does not allow us to address frequencies (in the spacecraft frame) higher than $\sim 100$Hz and considering that the Taylor-shifted electron gyroscale is typically $f_{\rho_e}\sim 70$Hz\footnote{For SW speed of $\sim 600$ km/s. Faster SW yields higher values of $f_{\rho_e}$ that can be larger than $100$Hz. In this case using STAFF-SA spectra may prove to be useful to study the electron dissipation range forming above $f_{\rho_e}$.} , STAFF-SA provides thus information only at a few frequencies in the interval [$8,100$]Hz which are $\sim$[$9,11,14,18,22,28,35,44,55,70,88$]Hz. This limited frequency resolution of the STAFF-SA spectra in the scale range between proton and electron gyroscales may result either in missing the spectral breaks or in not localizing them properly. In both cases determination of the scaling of the spectra at electron scales may be not very accurate. To show such a difficulty to locate the spectral break we plotted in Fig.\ref{sa} the same spectrum as in Fig.\ref{spectra}(c) (measured by STAFF-SC) as it would have been measured by STAFF-SA. We can see that while the spectral break can be observed easily on Fig.\ref{spectra}(c) (at $f\sim 30$Hz) it becomes more difficult to identify in Fig.\ref{sa}. The limited number of available points ($\sim 10$) would also make any fit subjet to high uncertainties (i.e. large error bars). 

This effect may be particularly relevant to explain the very low correlation ($C=0.03$) between the spectral breaks and $\rho_e$ reported in \citep{alexandrova12} as compared to the one shown in Fig.\ref{correl} ($C=0.72$), which is obtained using the STAFF-SC BM waveforms. STAFF-SA is certainly useful to provide information on the overall shape of the spectra at high frequency. However, it may not suffice to address specific details such as the actual scaling of the spectra at electron scales. When available, BM STAFF-SC data is certainly more appropriate to address this problem. The available range of frequencies ([$1,180$]Hz) is generally sufficient to address electron scale physics. In cases when electron scales would correspond to higher frequencies in the spacecraft frame (e.g. fast SW, or quasi-parallel whistler-like fluctuations) not accessible to STAFF-SC one then may use STAFF-SA spectra to probe into electron scales (provided that the SNR remains sufficiently high at those frequencies).

%------------------------------------------------------------------------
% Subsection III
%------------------------------------------------------------------------
\subsection{Effect of averaging the power spectra}\label{average}
\begin{figure}
\includegraphics[height=7cm,width=7.5cm]{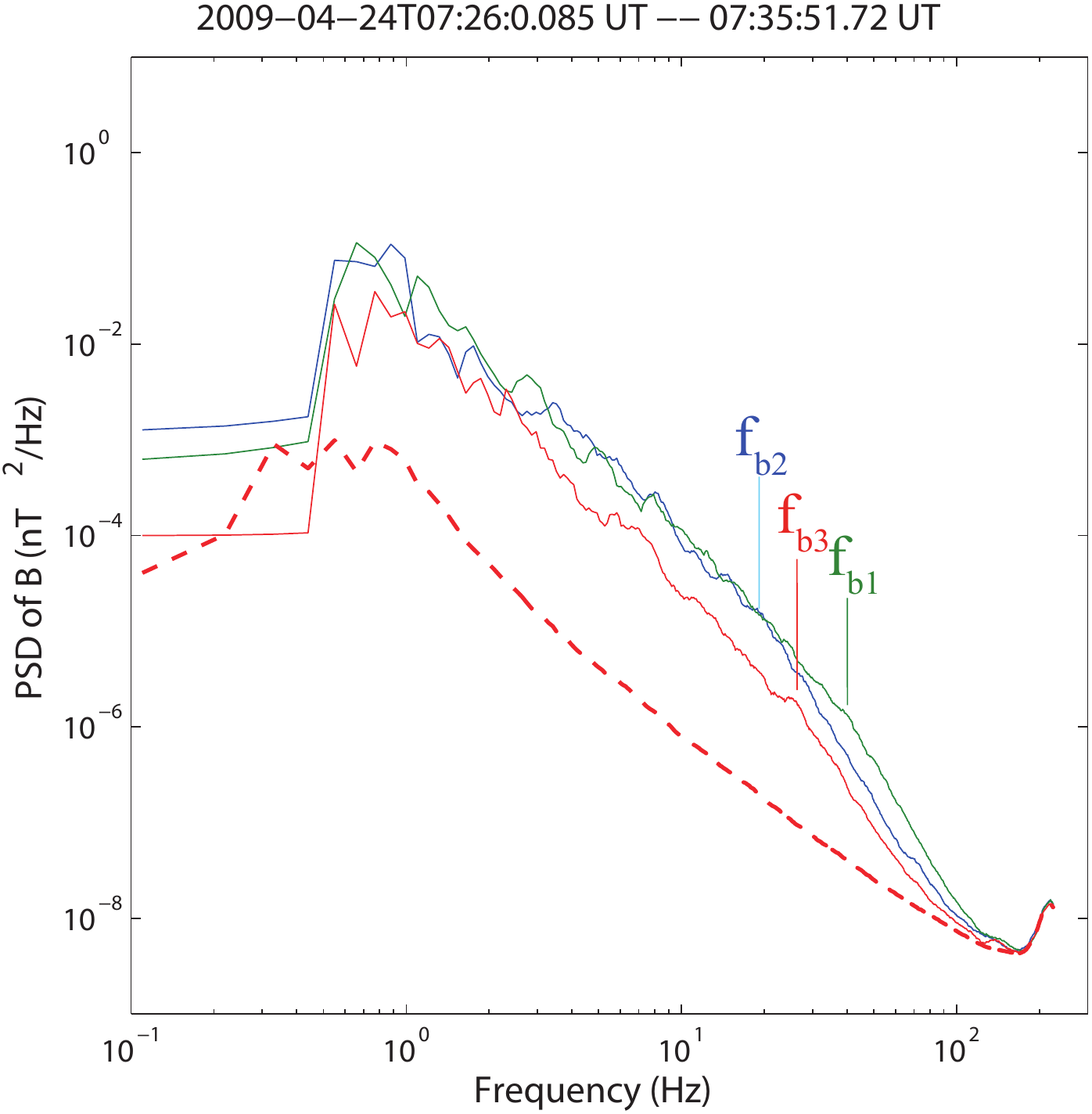}
\includegraphics[height=7cm,width=7.5cm]{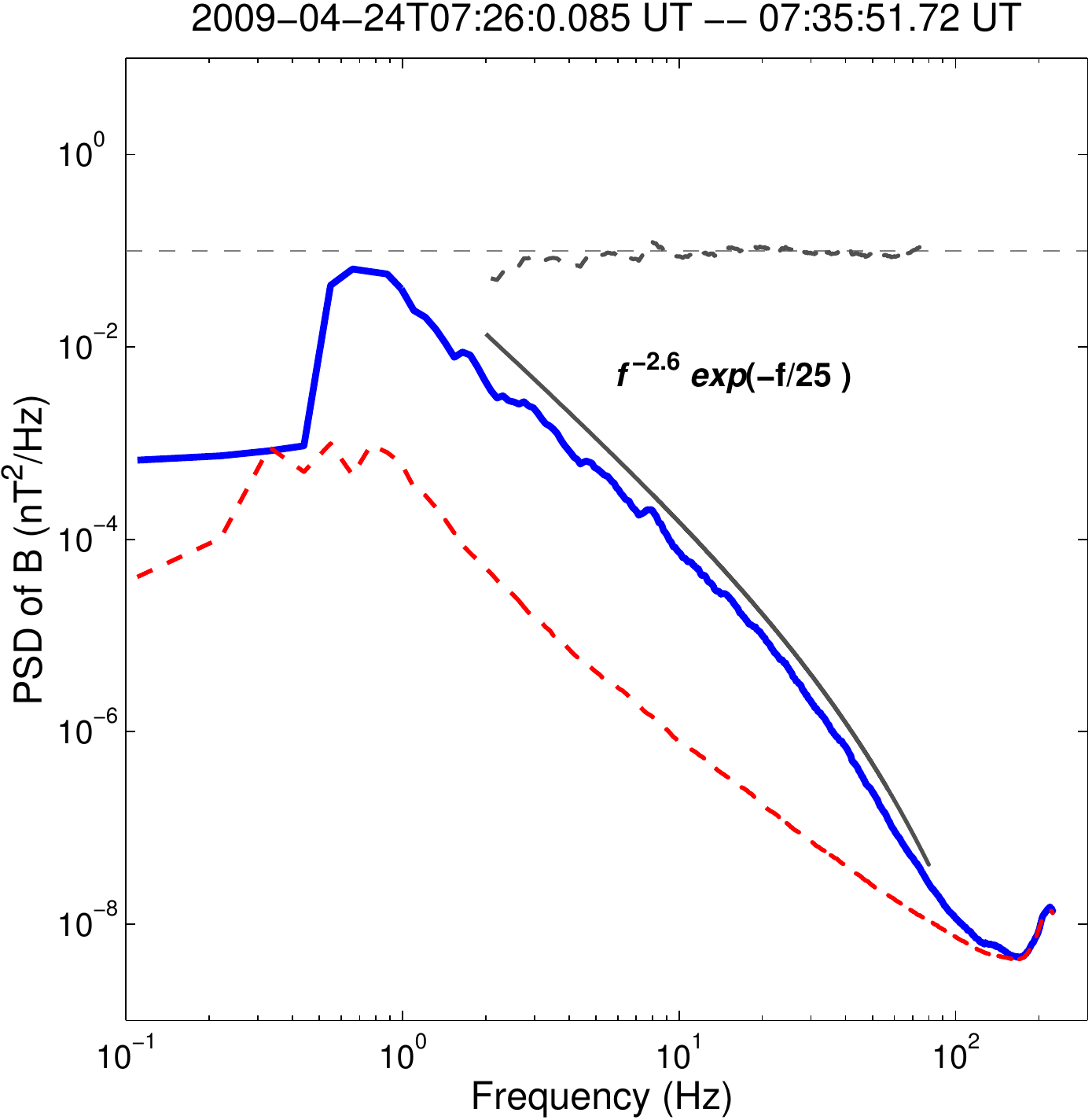}
\caption{Three magnetic energy spectra measured in the SW showing breakpoints at different frequencies indicated by the vertical lines (top). The  mean spectrum shows no clear break and curves at those frequencies (bottom). The horizontal curve (bottom panel) shows the compensated spectrum $B^2(f)/P(f)$ where $P(f)$ is the used fitting function.\label{spct_moy}}
\end{figure}

Another effect that may also influence the determination of the scaling of the turbulent spectra at electron scales is the  time (or frequency) averaging of the spectra, which is usually done to smooth the spectra under the assumption of time stationarity of the fluctuations. The approach consists generally in dividing a given time series into several shorter intervals of time for which a power spectrum is computed. Then the resulting spectra are averaged to reduce the fluctuations noise in particular at high frequencies. However, considering that the turbulent fluctuations are not stationary strictly speaking, averaging the spectra may result in missing some of the information contained in the individual spectra. This is particularly true regarding the presence of a spectral break at electron scales. For instance, let us assume that each given subinterval of time yields a power spectrum with a clear breakpoint at a given frequency $f_b$ in the spacecraft reference frame. If the value of the frequency $f_b$ changes from one subinterval of time to the other,  due for instance to fluctuations either in the mean SW parameters (e.g. $V_{sw}, T_i, T_e, V_A$) that enter in estimating the scale $\rho_e$ or the frequency $f_{\rho_e}$ corresponding to it and if the individual spectra have comparable power, then averaging them will result in smoothing the spectral breaks yielding a curved-like spectrum in the frequency range corresponding to all the breaks. To show this we plotted in Fig. \ref{spct_moy} (top panel) three spectra computed over $10$sec within a window of $10$mn. The three spectra show clear breakpoints approximately at $20$Hz,  $30$Hz and $45$Hz. When the three spectra were averaged (bottom panel) they yield a spectrum without any clear break in the frequency range [$20,45$]Hz. The spectrum at those frequencies curves and can be well fit by exponential-like models.

%-----------------------------------------------------------------------
% Subsection IV
%-----------------------------------------------------------------------
\subsection{Effect of instrumental interferences at high frequency}\label{average}
\begin{figure}
\includegraphics[height=5cm,width=7cm]{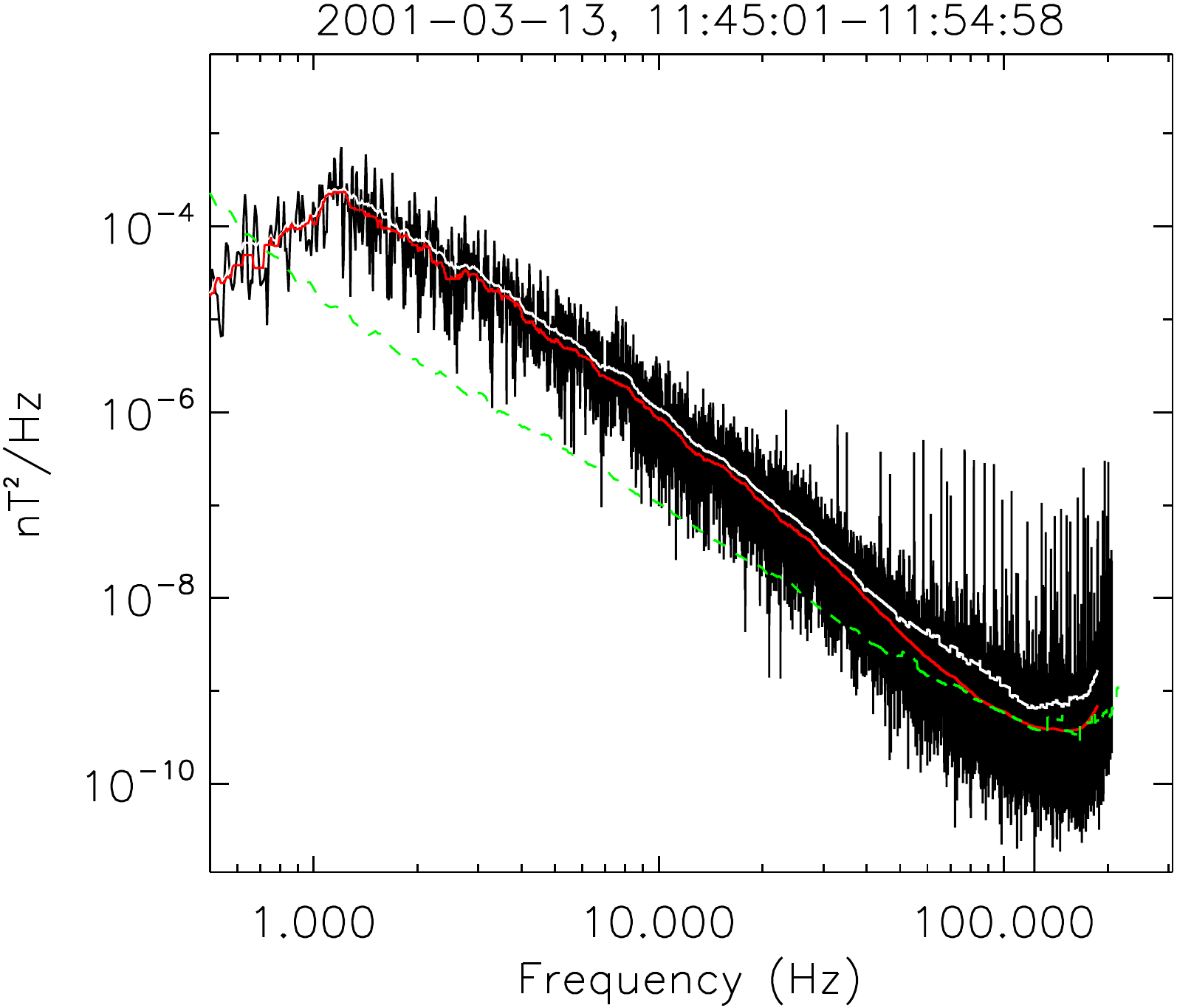}
\includegraphics[height=5cm,width=7cm]{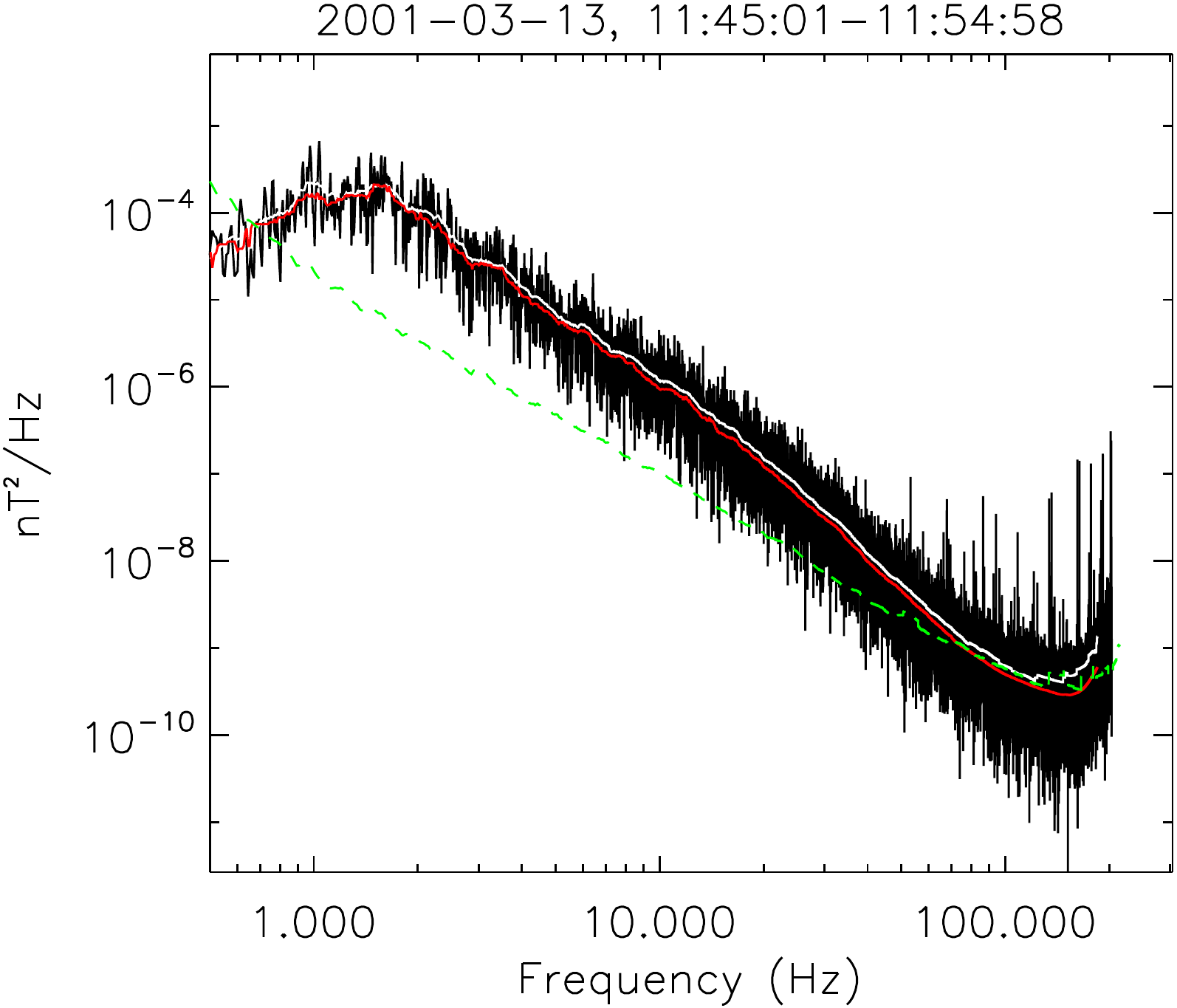}
\caption{An example of high frequency spikes observed on the spectra of $B_z$ measured by the STAFF instrument which may be caused by interferences coming from the DWP instrument onboard spacecraft 1 (top) and spacecraft 2 (bottom). The corresponding smoothed spectra by a mean (white) and the median (red) functions are shown. The dashed green line is the in-flight mean sensitivity floor of the STAFF instrument. \label{dwp}}
\end{figure}

The last instrumental problem that can affect the proper determination of the scaling the magnetic energy spectra in the SW is the presence of spectral interferences observed at high frequency of STAFF-SC spectra. These interferences may be caused by the Digital Wave Processor (DWP) onboard Cluster. The DWP instrument is designed to synchronize the wave instruments of Cluster by emitting a signal allowing communication between  them. This signal might be at the origin of the interferences with the STAFF magnetometers (similar interferences are also observed on FGM spectra). Examples of such interferences are shown in Fig. \ref{dwp}. 

Although the spikes caused by such interferences are few, their relatively high amplitudes may nevertheless affect the fit of the turbulent spectra at electron scales. Indeed, when smoothed the intense spikes lead to over estimating the power at the highest part of the spectra, which then produces shallower spectra as shown in Fig. \ref{dwp} (top). The figure compares smoothings by a mean and a median functions on windows of size logarithmically increasing with frequency. The effect of interferences becomes more important for spectra with low SNR, which emphasizes the need to  use the highest available SNR in the SW. 

From Fig.\ref{dwp} we can also see that the four Cluster satellites are not affected equally by the interferences. A more complete study (not shown here) revealed that spacecraft 1, and particularly its $B_z$  component, is most affected. Therefore, considering that most of the studies of the electron scales in the SW are based on single spacecraft data we recommend to use data from the spacecraft $2$, $3$ or $4$. Depending on the level of the SNR, the number and amplitudes of the spikes that would appear on the spectra, the corresponding data may need to be pre-processed by removing the spikes prior to any analysis. 

Finally, we note that these interferences would also affect the spectra computed by the STAFF-SA instrument onboard the spacecraft. Owing to the fact that the STAFF-SA spectra are averaged over a frequency band centered at each of the $27$ frequencies, the contribution of the spikes may be accounted for in the transmitted energy density. Since the original waveforms are lost, it is not possible to quantify the effect of the interferences on the STAFF-SA spectra.

%---------------------------------------------------------------------------------------------------------------------------------------------------------------------------
%%%%%%%%%%%%%%%%%			Nouvelle section             %%%%%%%%%%%%%%%%%%
%---------------------------------------------------------------------------------------------------------------------------------------------------------------------------
\section{Data selection}\label{data}
\begin{figure}
\includegraphics[height=4.5cm,width=8.5cm]{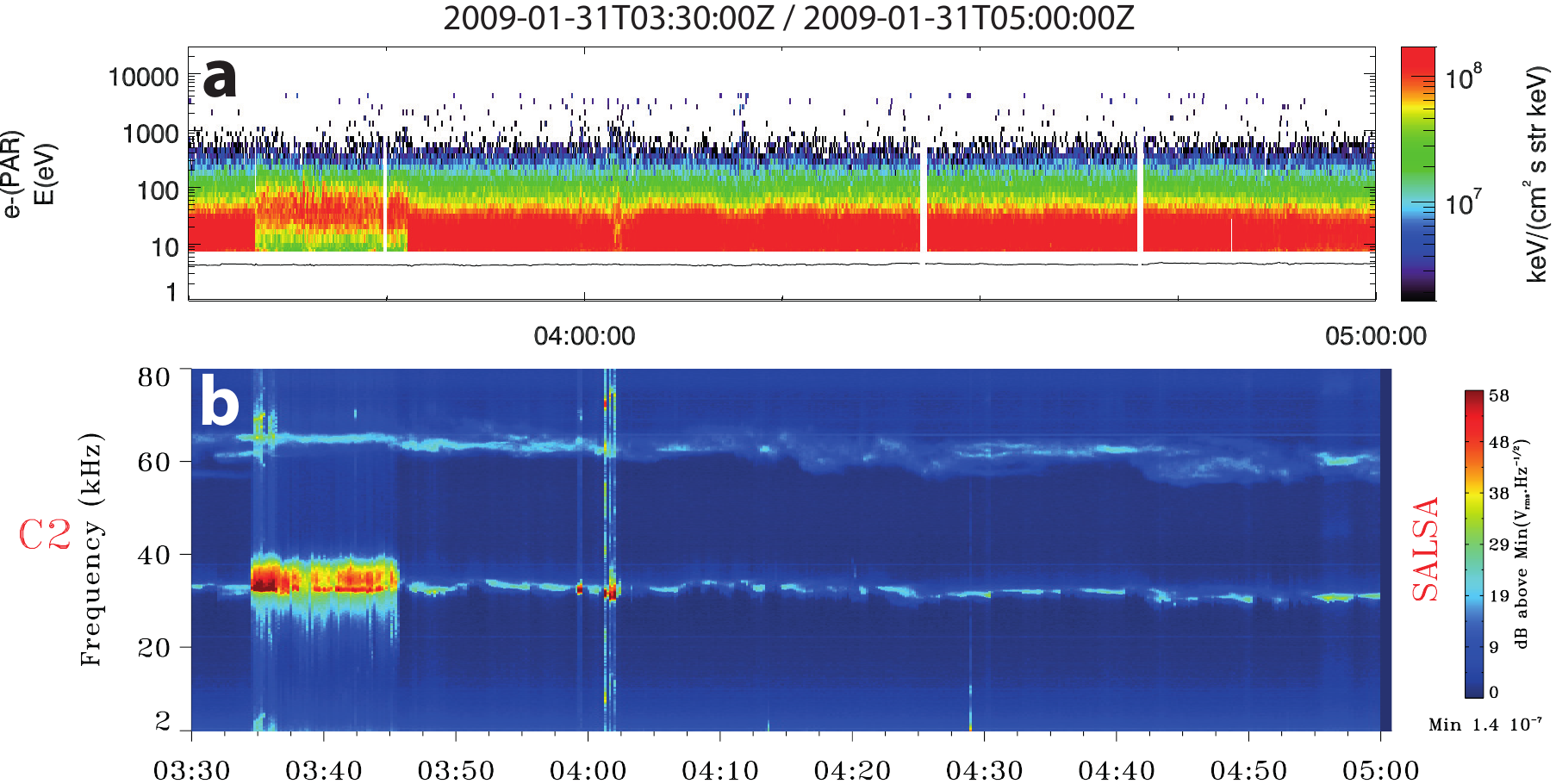}
\caption{A typical example of the analyzed events. (a) is the spectrogram of field-aligned electron energy and (b) is the electric field from PEACE (SC1, no data available on SC2) and WHISPER (SC2) showing that the spacecraft are in the free SW from 04h05 to 05h00 (no foreshock electrons). \label{whisper}} 
\end{figure}

In the present study we surveyed all STAFF-SC BM data measured  in the SW by the Cluster 3 spacecraft from $2000$ to $2011$. We used AMDA data base  (http://cdpp-amda.cesr.fr/DDHTML/index.html) to select intervals of time when the Cluster spacecraft was located in the SW. AMDA provides useful routines to search for data according to desired physical parameters (e.g., high or low SW speeds, plasma $\beta$). The obtained list of events is then checked event by event with respect to WHISPER and PEACE data to remove any interval containing foreshock electrons. An example of the targeted SW intervals is shown in Fig.\ref{whisper}.

To eliminate data from the ion foreshock region (IFS), or from other boundaries such as Interplanetary Shocks (IPS) or CMEs requires additionally to use data from the Cluster Ion Spectrometer (CIS). The free SW is indeed generally characterized by a beam of ions with an energy around $1-2$ KeV. Any enlargement of the beam toward higher energies generally indicates a heating of ions across boundaries such as shocks. The CIS spectrograms available at the ESA Cluster Active Archive (CAA) web site (http://caa.estec.esa.int/caa/home.xml) allow one to identify the IFSs and IPSs. However, the spectrograms cannot guarantee the absence of any reflected ions from the shock, which can only evidenced by a careful analysis of the Ion Distributions Functions (IDFs). In the following, to eliminate the significant EFSs, IFSs and IPSs we rely mostly on the examination of the Whisper, PEACE and CIS spectrograms. If afterwards a doubt still remains we then analyze the corresponding IDFs provided at the CAA web site to check the presence of reflected energetic ions from the foreshock\footnote{The IDFs given in the ESA/CAA website may need a careful post-treatment to obtain clean IDFs in the SW. However, doing such detailed studies of the IDFs is hardly feasible for all our events~\citep{mazelle12}.}. Despite these numerous checks of the data one still needs to assume that if any residual reflected ions from the foreshock would still exist in our data (not observable on the spectrograms or in the analyzed IDFs) they do not have significant impact on the electron scale physics, which is the main purpose of the present paper. Indeed, such ions when they exist would affect ion scales $\sim \rho_i$ (rather than electron scales $\sim \rho_e$) through, for instance, ion plasma instabilities that might be generated in the SW plasma. 

\begin{figure}
\includegraphics[height=5cm,width=7.5cm]{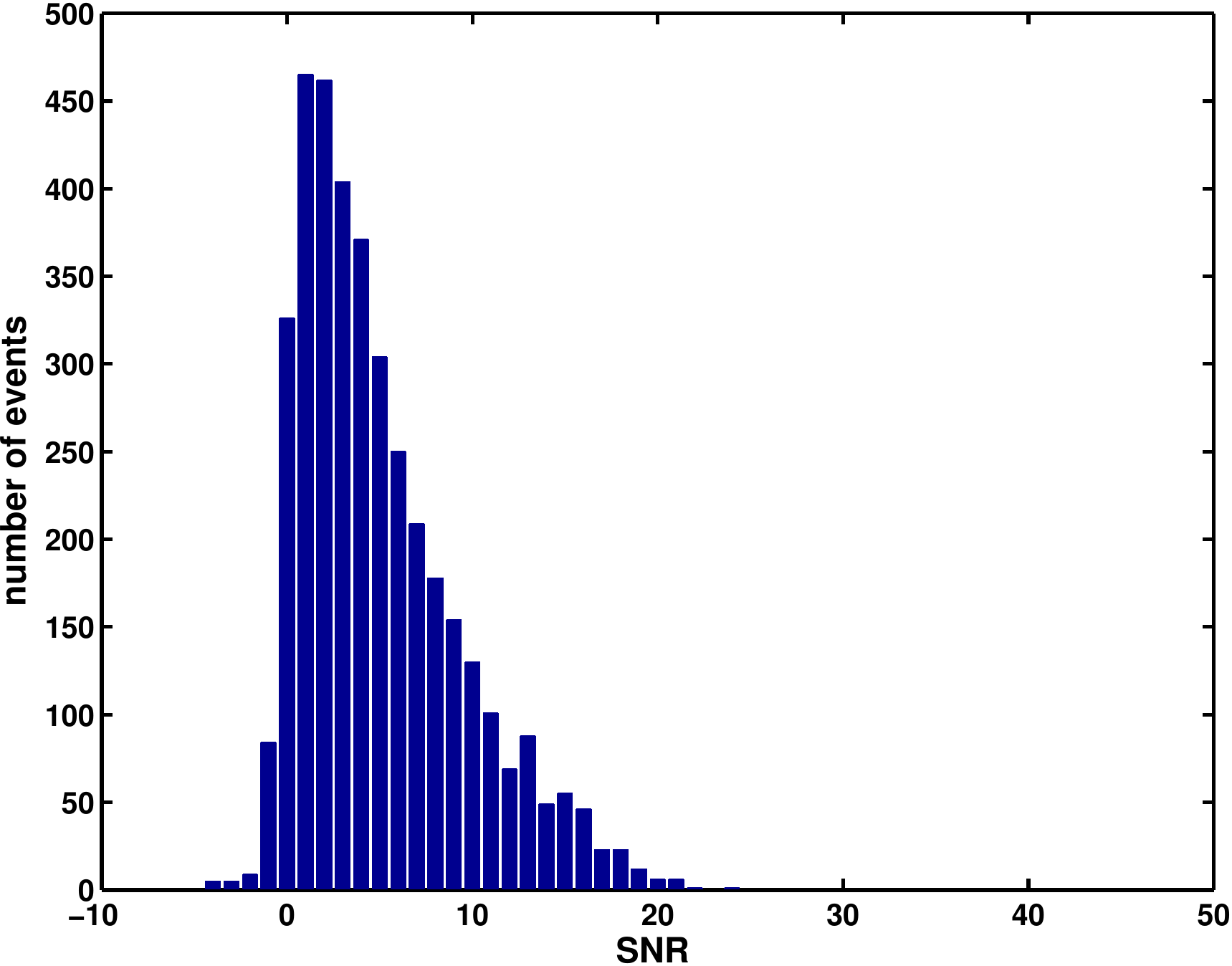}
\caption{The distribution of the Signal-to-Noise-Ratio (defined in the text) of the $B_z$ magnetic energy spectra at $f=30$Hz. \label{snr}}
\end{figure}

After obtaining the list of events that survive the different criteria given above, we focus only on intervals where magnetic fluctuations have a high SNR. To do that we computed the magnetic energy spectra over approximately $10$s for all intervals and estimated the value of the SNR at the frequency $f_{sc}\simeq30Hz$. With a SW speed $V_{sw}\sim 500$km/s and an electron gyroradius $\rho_e \sim 1$km, the frequency $f_{sc}=30$Hz corresponds roughly to the scale $k\rho_e\sim 0.4$ using the Taylor assumption (other histograms of SNR were computed at neighboring frequencies and found to be very similar to the one shown here). The obtained histogram (Fig.~\ref{snr}) shows that at $f_{sc}\simeq30$Hz most of the turbulent spectra reach the sensitivity floor of the Cluster SCM, which means that those events cannot be used to address the electron scale physics. To achieve this goal we need to select intervals with the highest possible SNR. However, as can be seen on Fig.~\ref{snr}, the higher is the SNR the rare are the events. Here we present only the results for events having SNR$\geq10$, and a number of $620$ was found. These events represented about $15\%$ of the total events (an ``event" refers to a total power spectrum computed over nearly $10$s period of time).

%---------------------------------------------------------------------------------------------------------------------------------------------------------------------------
%%%%%%%%%%%%%%%%%			Nouvelle section             %%%%%%%%%%%%%%%%%%
%---------------------------------------------------------------------------------------------------------------------------------------------------------------------------
\section{Fitting models}\label{fits}
As we discussed above, the theoretical and numerical studies of SW turbulence dissipation at electron scale are very recent. The absence of firm theoretical predictions on this problem and the limitations of the available data make it difficult to draw firm conclusions on the actual processes of dissipation, or even on the actual scaling of the energy spectra at electron scales. To illustrate that we plotted in Fig.\ref{fits} an example of a magnetic energy spectrum measured in the SW by the STAFF-SC experiment (black curve). We also over plotted different models to fit the data:
\begin{itemize}
\item {\it Double-power-law model}
\begin{displaymath}
P(f)=A_1f^{-\alpha_1}\big[1-H(f-f_b)\big]+A_2f^{-\alpha_2}H(f-f_b)
\end{displaymath}
\item{{\it Exponential model}}
\begin{displaymath}
P(f)=Aexp(-a\sqrt{f/f_b})
\end{displaymath}
\item{Hybrid model}
\begin{displaymath}
P(f)=Af^{-\alpha_1} exp(-f/f_b)
\end{displaymath}
\item{Asymptotic-double-power-law model}
\begin{displaymath}
P(f)=A\frac{f^{-\alpha_1}}{1+(f/f_b)^2}
\end{displaymath}
\end{itemize}
Above $H(f)$ is the Heaviside function, $f_b\sim 25$Hz, $\alpha_1=2.6$ and  $\alpha_2=4.2$, and $a=15$.
The exponential model has been proposed in \cite{alexandrova09} to fit the magnetic energy spectra in the scale range $\sim[0.1,1]k\rho_e$\footnote{To change from frequency to wavenumber one can replace $f/f_b$ by $k\rho_e$ using the Taylor hypothesis and assuming $f_{\rho_e}\sim f_b$.}. The hybrid model is reported in \cite{alexandrova12} and combines the two  previous ones where the power-law dominates the lowest part of the spectrum while the exponential function starts acting as frequencies approach $f_b$ yielding a curved-like spectrum. The last model is asymptotically similar to the double-power-law model: for $f<<f_b$ $P(f)\sim Af^{-\alpha_1}$ and for $f>>f_b$ $P(f)\sim f^{-\alpha_1-2}\sim f^{-4.6}$. However, when $f\sim f_b$ this model is very close to the hybrid model and results in a curved-like spectrum.

One should stress here that these functions are chosen for pure mathematical reasons and they do not necessarily reflect  actual physical mechanisms of dissipation, although there may exist a possible connection between a given scaling with a particular dissipation process. For instance hydrodynamic turbulence is known to yield an exponential dissipation, while asymptotic-double-power-law models have been reported in fusion plasma turbulence where the Finite Larmor Effects (FLRs) play a key role \citep{hasegawa78,gurcan09}. 

\begin{figure}
\includegraphics[height=6.5cm,width=8cm]{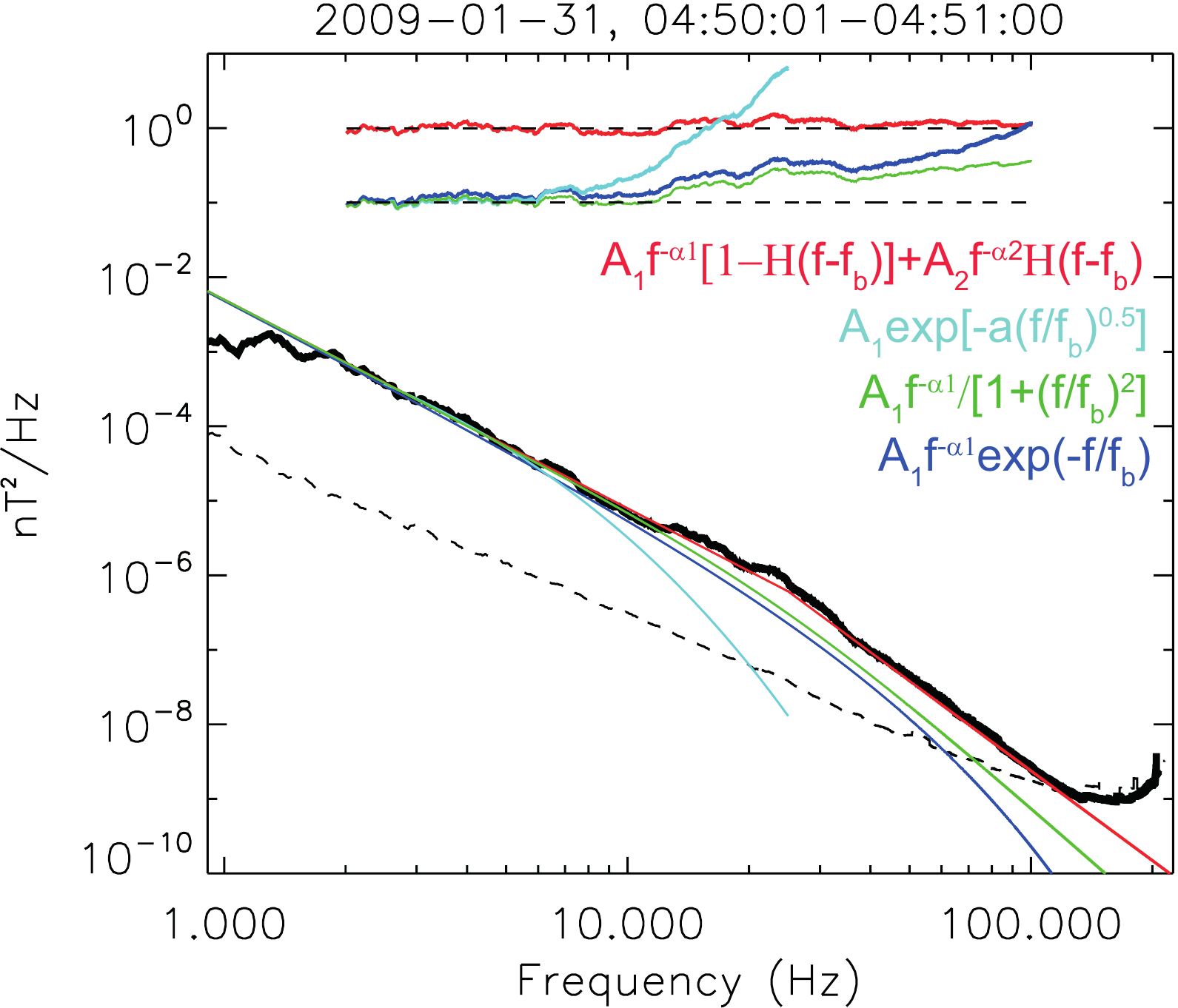}
\caption{Example of a magnetic energy spectrum of $B$ measured in the SW by the Cluster 2/STAFF-SC instrument (black). The dashed line indicates the estimated (in-flight) sensitivity floor of the instrument. Examples of fitting functions are plotted for comparison:  double-power-law (red), asymptotic-double-power-law (green), hybrid (blue) and exponential (cyan). The horizontal curves are compensated spectra $B^2(f)/P(f)$, where $P(f)$ is the corresponding fitting function.\label{fits}}
\end{figure}

The comparison between the fits in Fig.\ref{fits} shows indeed that the double-power-law model fits better the data as it captures the observed spectral break. The exponential model is shown not to  reproduce correctly the scaling in the range $k\rho_e \sim [0.1,1]$ as reported in \cite{alexandrova09}. This model will not be considered in the rest of the paper. However, as expected, the hybrid and the asymptotic-double-power-law fail to fit the spectral break and the neighboring frequencies but they reproduce the overall shape of the spectrum, in particular the asymptotic-double-power-law. Nevertheless, these two models are very close to each other, significant differences appear indeed only at high frequency where the noise floor of the instrument is reached. From this one concludes that, given that the highest SNR values available from the Cluster/SCM data in the SW are at best similar to that of Fig.~\ref{fits}, it is unrealistic to hope distinguishing between these two models in  SW observations. However, the double-power-law model can be distinguished from the three other (curved) models provided that the spectral break can be evidenced properly. As discussed above to detect the spectral break at electron scales requires particular caution as its presence may easily be smoothed by several effects.  

%---------------------------------------------------------------------------------------------------------------------------------------------------------------------------
%%%%%%%%%%%%%%%%%			Nouvelle section             %%%%%%%%%%%%%%%%%%
%---------------------------------------------------------------------------------------------------------------------------------------------------------------------------
\section{Results}\label{results}
\begin{figure}
\includegraphics[height=7.5cm,width=8.5cm]{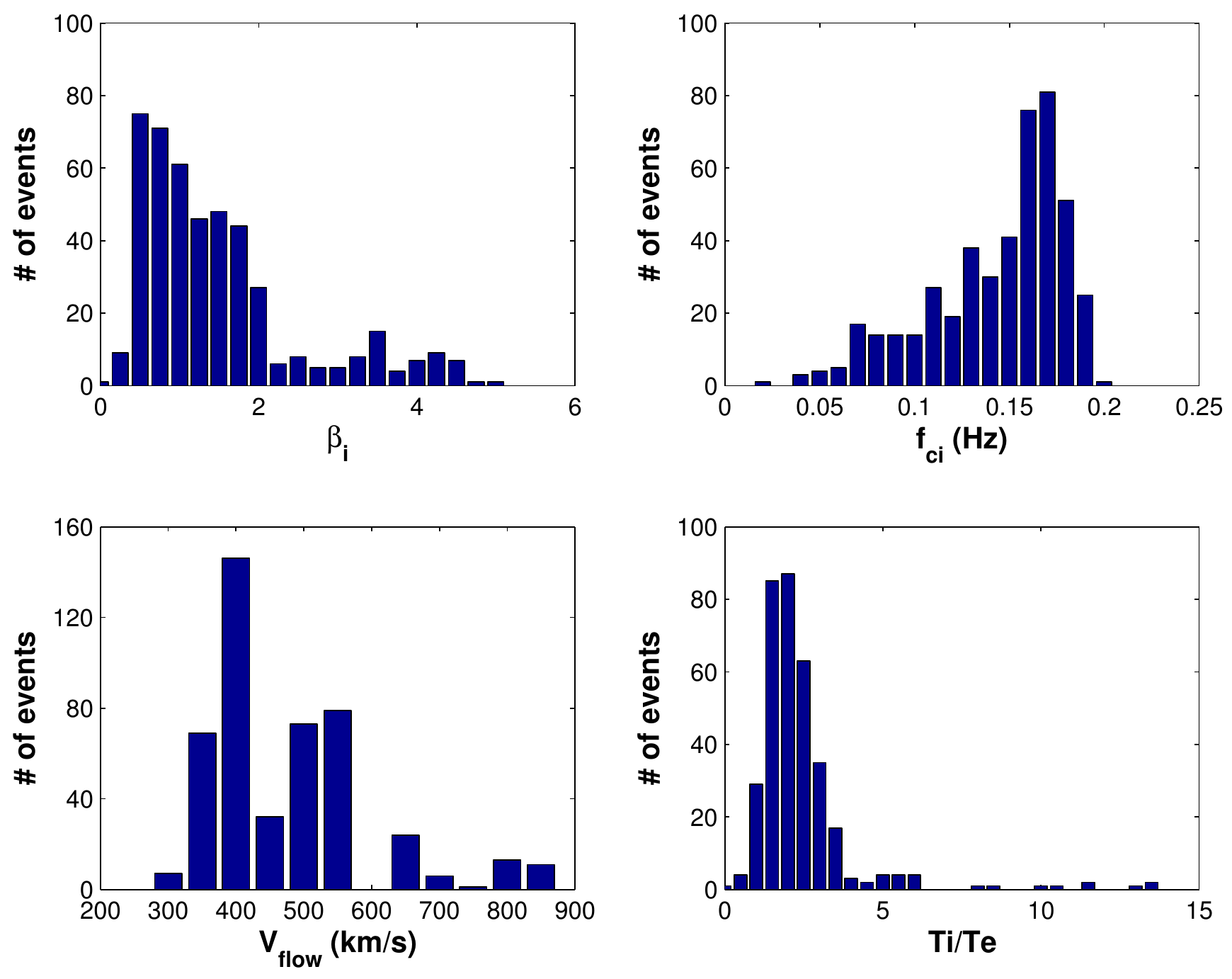}
\caption{The histograms of different mean SW parameters: $\beta_i$, ion cyclotron frequency $f_{ci}$, the solar wind speed $V_{flow}$ and the ratio of ion to electron temperatures. \label{parameters}}
\end{figure}

For the obtained list of events we computed the histrograms of the corresponding mean plasmas parameters, which are given in Fig.~\ref{parameters}. We found in particular that most of the events have $T_i>T_e$ and $\beta_i \gtrsim1$. For these events we computed the corresponding power spectra of the magnetic fluctuations over $10s$. This time length has been chosen to allow for covering the frequency range $\sim$[$1,180$]Hz. To study lower frequencies longer time intervals are required, but these frequencies will not be addressed here. 

\begin{figure}
\includegraphics[height=6cm,width=7.5cm]{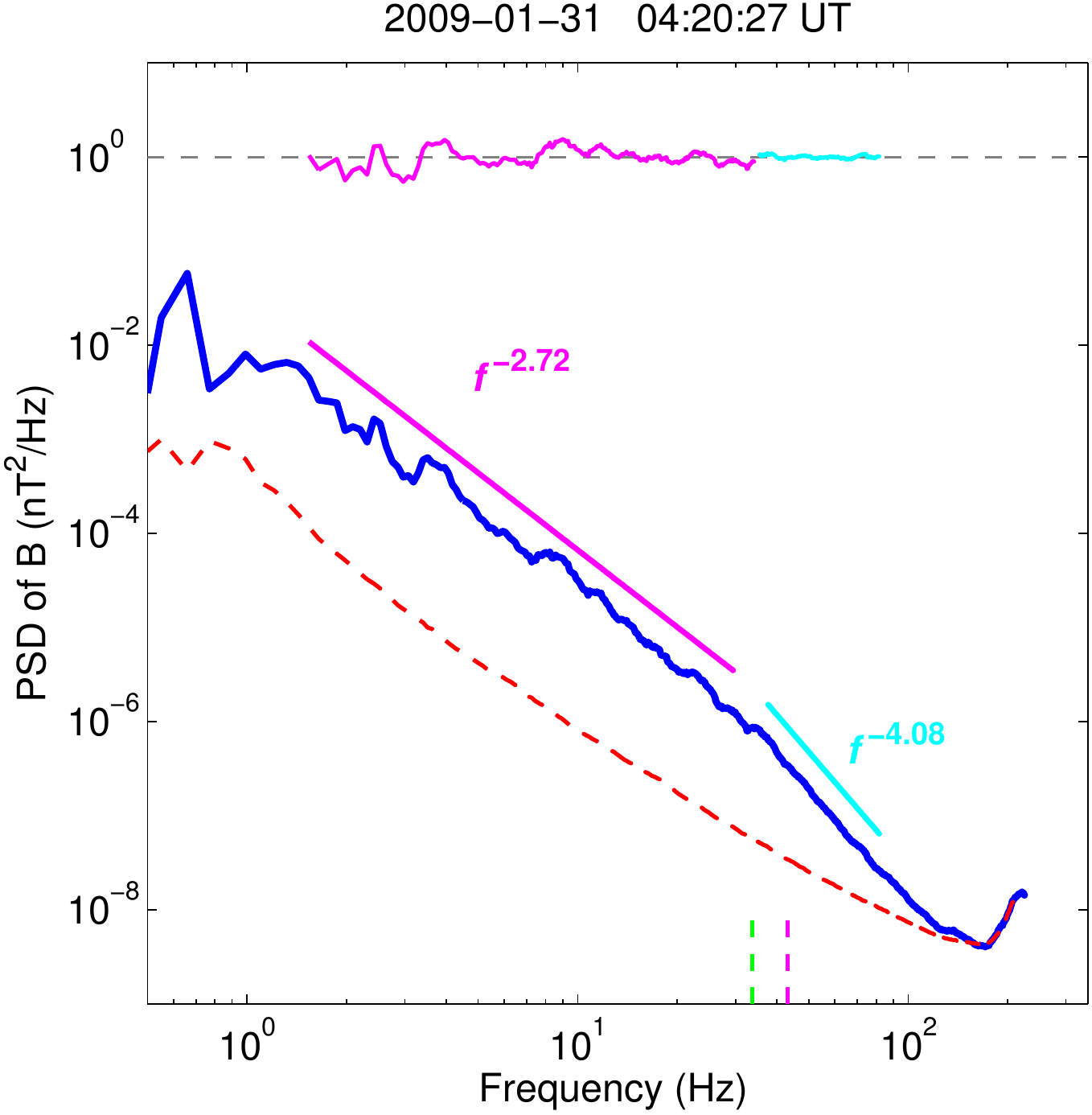}
\includegraphics[height=6cm,width=7.5cm]{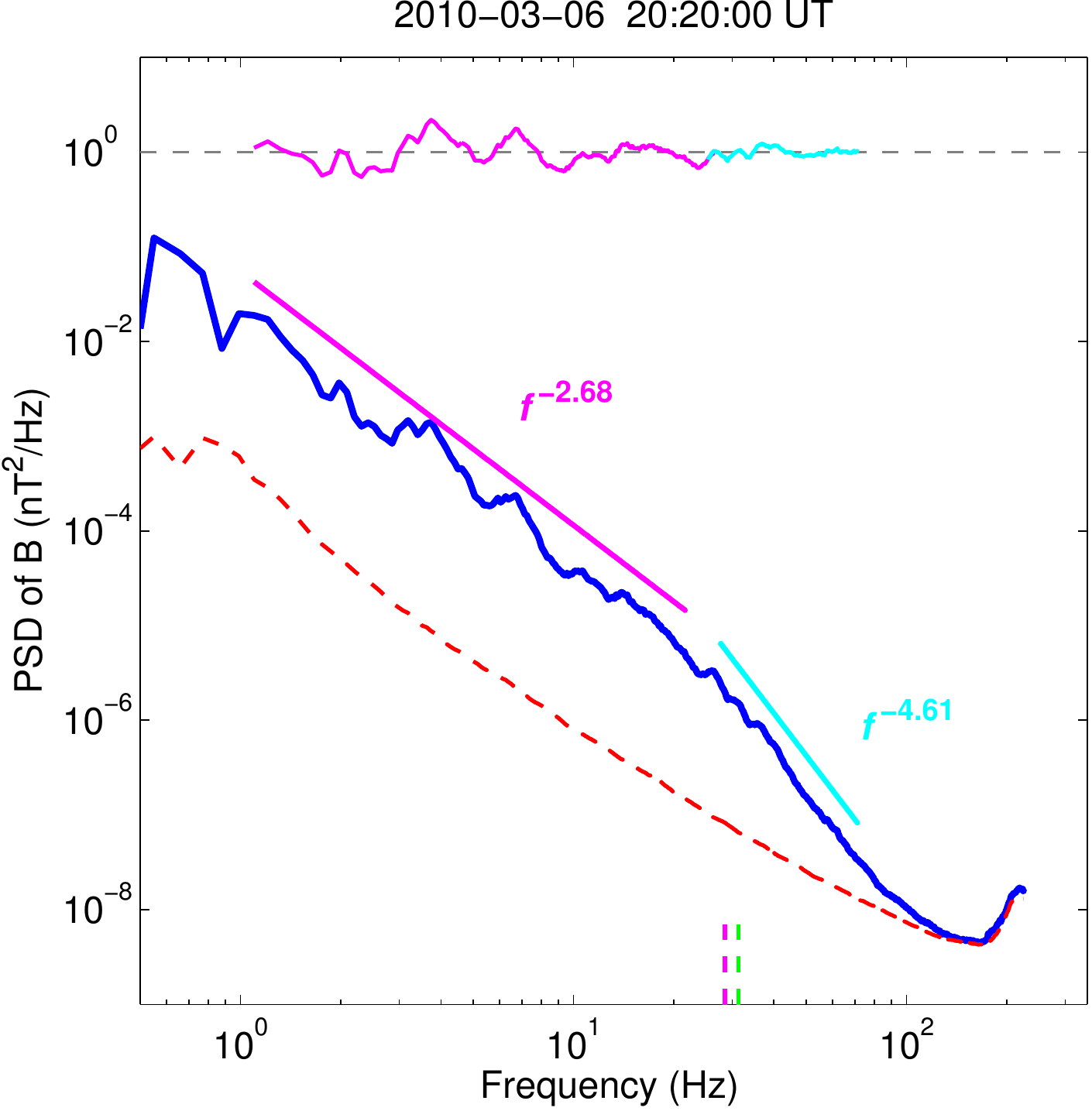}
\caption{Examples of the magnetic energy spectra measured in the free SW showing clear spectral break at electron scales (vertical dashed lines). A similar description as in Fig.\ref{fits} applies.\label{spectra}}
\end{figure}

Some examples of the obtained spectra are plotted on Fig.~\ref{spectra}. They show that the magnetic energy cascades below the ion scale and reaches the electron gyroscale where a clear break is evidenced, in agreement with earlier observations reported in \cite{sahraoui09, sahraoui10a}. Similarly to Fig.~\ref{fits}, we plotted on Fig.\ref{spectra}(a) the three previous fitting models to check which one fits better the spectra. As can be seen on the compensated spectra plotted on the top of the figure, the double-power-law is found to fit better the observed spectra at the scales near and below the spectral break. The presence of a clear break near the electron scales invalidates the argument that the break at electron scales is essentially caused by the foreshock electrons present in the data analyzed in \cite{sahraoui09}. Indeed, the presence of foreshock electrons may generate a ``bump" in the magnetic spectra at that scale as shown in Fig.3(a) in \cite{sahraoui09}, but it is not clear as to how they can generate a spectral break followed by a steep power-law spectrum as shown in Fig.3(b) of \cite{sahraoui09}. In any case, the results shown here, which were obtained from SW data that contained no foreshock electrons clearly prove that the spectral break cannot be attributed primarily to foreshock  electron. It most likely results from the nonlinear dynamics of the plasma itself. This indeed has been observed in 2D and 3D PIC simulations \citep{camporeale11, chang11}, predicted by existing theories on SW  turbulence~\citep{schekochihin09,meyrand10} and observed also in magnetosheath turbulence~\citep{huang13}. The disagreement with the observations of \cite{alexandrova09} can be explained by the lack of universality of turbulence at electron scales, or by the caveats discussed in section \ref{caveats}. Note however that the recent study by \cite{alexandrova12} showed also that $30\%$ (out of $100$) of the analyzed STAFF-SA spectra agreed with the double-power-law (or the break) model discussed here.

\begin{figure}
\includegraphics[height=4.5cm,width=8cm]{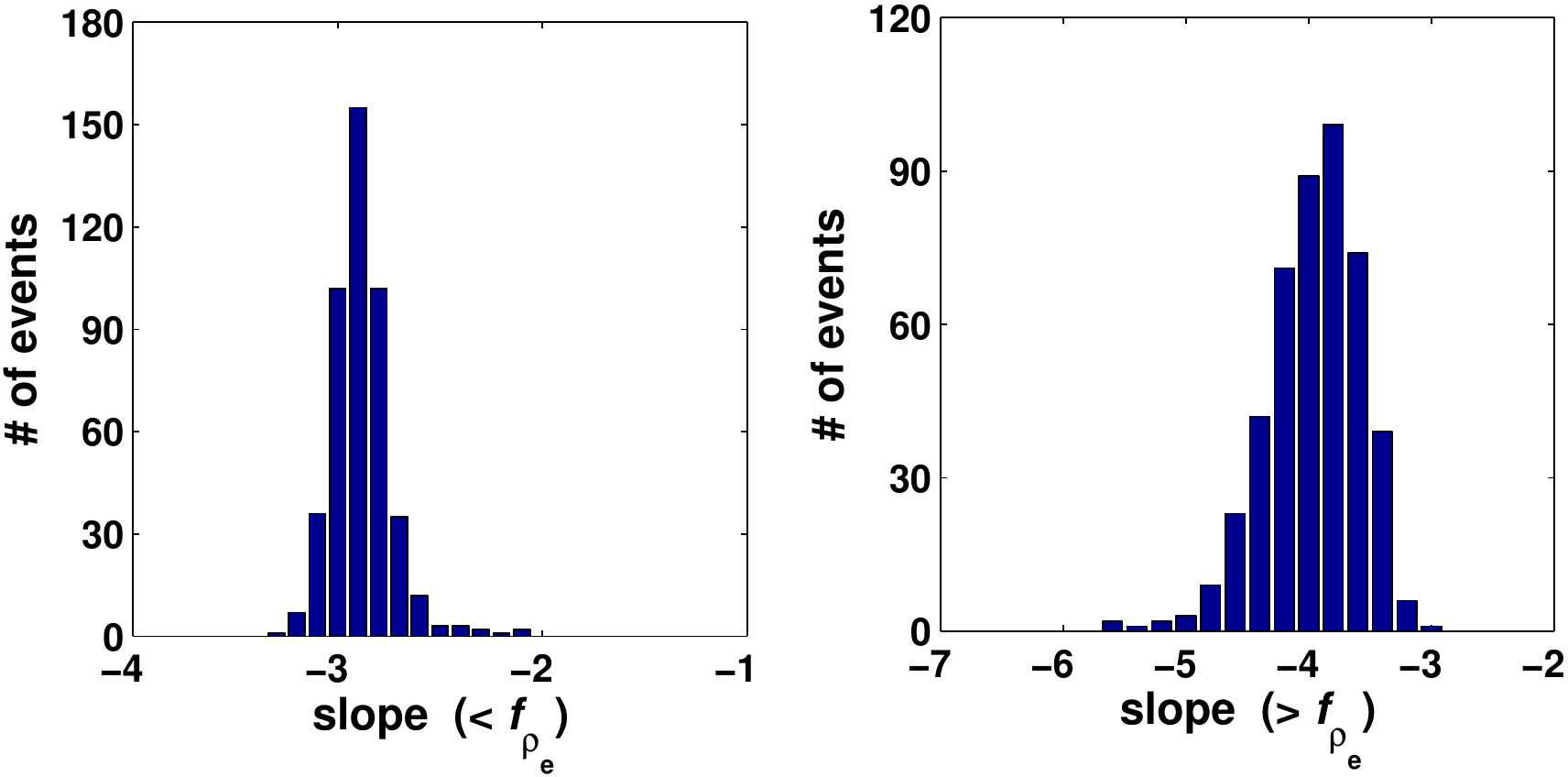}
\caption{The histograms of the slopes resulting from a double-power-law fits of the scales below (left) and above (right) the frequency $f_b$ corresponding to the spectral break near the electron gyroscale.\label{slopes}}
\end{figure}

Following the arguments given here we performed similar double-power-law fits below and above the spectral break to all events found to have breaks (some events, mostly those corresponding to fast SW, do not show breaks in the frequency range accessible to STAFF-SC). The histograms of the obtained slopes are given in Fig.~\ref{slopes}. We can see that the slopes of the spectra in the dispersive range (i.e. [$f_{\rho_i}$, $f_{\rho_e}$]) cover the range  $\sim[-2.5,-3.1]$ with a peak at $\sim -2.8$. These values are close to those predicted by existing theoretical and numerical simulations \citep{camporeale11, chang11, howes11a, boldyrev12} and to the observations reported in \cite{alexandrova12}. The spectra below $f_{\rho_e}$ are however steeper with slopes distributed in the range $\sim [-3.5, -5.5]$ and a peak at $\sim -4$, which are in general agreement with the predictions from the 2D and the 3D PIC simulations \citep{camporeale11,chang11} and with the results of the break model reported in \cite{alexandrova12}. This wide distribution suggests the lack of universality of turbulence at electron scales as compared to that in the inertial range. A similar study has been performed on magnetosheath turbulence where steeper spectra above $f_{\rho_e}$ were observed whose slopes distributed in the interval $\sim [-4, -7.5]$ and a peak near $-5.5$ \citep{huang13}. We will discuss this point further below. 

Another important question that we can try to answer here is the actual physical scale, $\rho_e$ or $d_e$, that would correspond to the spectral breaks observed within the data. This question relates to the appropriate theoretical description of the electron scales physics~\citep{schekochihin09,bourouaine12}. In the so-called Hall-MHD approximation, which is valid in the limit of $\beta_i<<1$ and $T_i<<T_e$, the Hall term brings a correction to the classical MHD theory proportional to $d_i$, the ion inertial length. This theory can be extended to electron scale physics within the incompressible Electron-MHD (EMHD)~\citep{biskamp99,meyrand10} or the reduced two-fluid-theory \citep{sahraoui03b,sahraoui12}. Within the EMHD model, which is appropriate to describe whistler mode turbulence, $d_e$ arises as the relevant scale to describe electron scales physics. However, in the SW at $1$AU, where $\beta_i\sim1$ and $T_i/T_e\gtrsim 1$ (see Fig.\ref{parameters}), these popular models are not rigorously valid \citep{schekochihin09,howes09}, although they may reproduce some observed properties of SW turbulence such as the scaling of the magnetic or the electric energy spectra in the dispersive range \citep{matthaeus08}. 

In kinetic theory, 3D recent PIC simulations of electron scales whistler turbulence showed that  a spectral break occurs near $d_e$ \citep{chang11} and not at $\rho_e$ (note that in that paper $\beta_e=0.1$, which means that $\rho_e=d_e/\sqrt{\beta_e} \sim 3d_e$). On the other hand kinetic theory of KAW turbulence in the limit $k_\parallel <<k_\perp$ predicts that the relevant scales are $\rho_i$ and $\rho_e$. A simple explanation of this can be given by examination of the linear dispersion relations of the KAW at different values of $\beta_i$ (or $\beta_e$) shown in Fig. \ref{disp} (top). One can see that regardless the value of $\beta_i$ the KAW becomes dispersive at $k\rho_i\sim1$ and no significant change is observed at $kd_i\sim 1$.

\begin{figure}
\includegraphics[height=4cm,width=8.5cm]{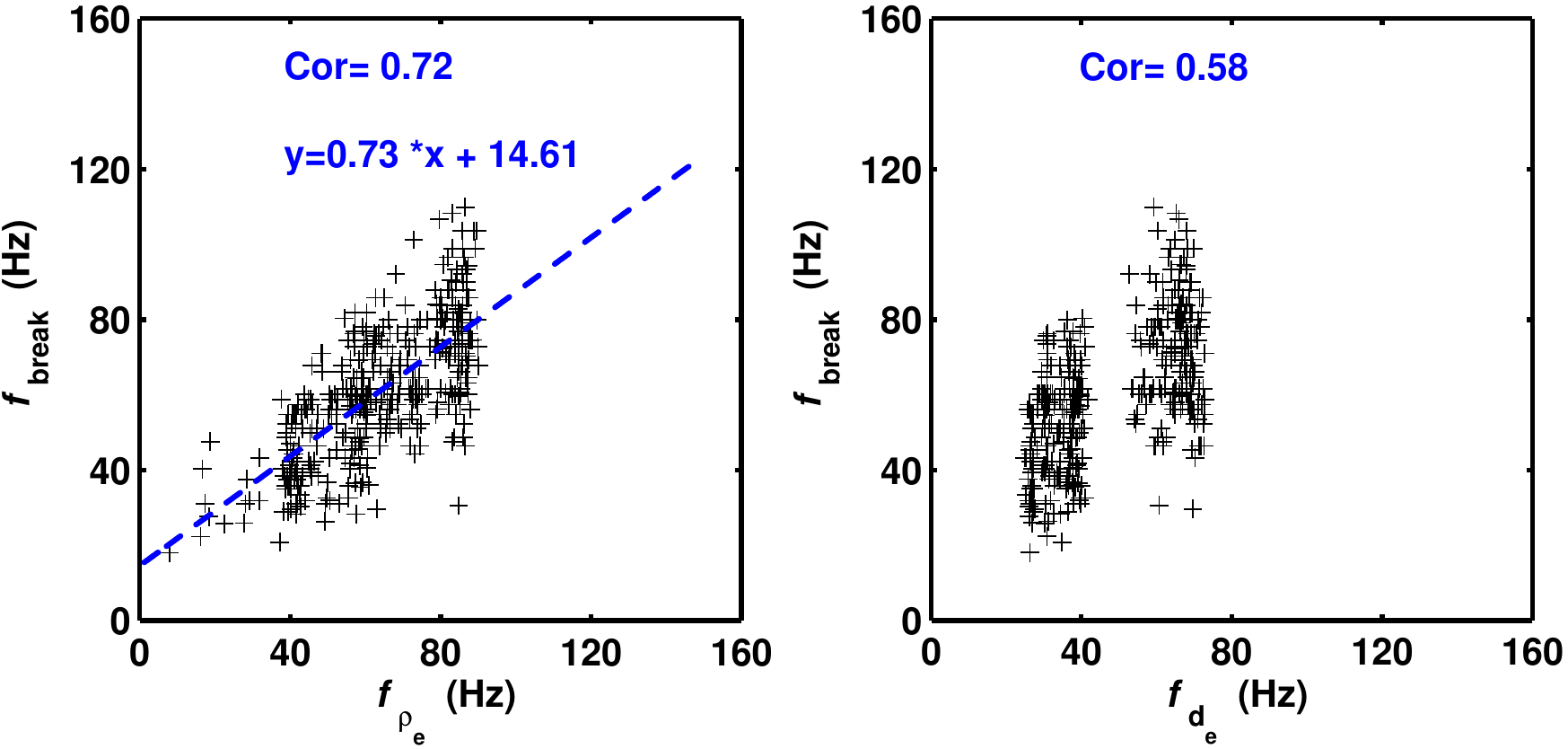}
\caption{Correlation between the frequencies of the observed spectral breaks $f_b$ with the Taylor-shifted electron gyroscale $\rho_e$ (right) and inertial length $d_e$ (left).\label{correl}}
\end{figure}

From the point of view of observations answering the question as to whether $\rho_i$ or $d_i$ is the relevant scale is more difficult because, often, SW data at $1$AU show that $\beta_i\sim 1$, which means that one cannot distinguish the two scales. Nevertheless, some statistical studies have tried to answer this question~\citep{leamon99,bourouaine12}. For electron scales only a few studies based on a limited number of events have reported the relevance of $\rho_e$ to characterize electron scale physics in the SW \citep{sahraoui09,alexandrova09,sahraoui10a}. \cite{alexandrova12} have confirmed that result over a larger data samples ($100$ spectra). In Fig.~\ref{correl} we computed the correlation between the frequencies of the observed spectral breaks with $f_{\rho_e}$ and $f_{d_e}$. The figure shows a relatively high correlation (0.72) of the spectral break with the electron gyroscale $\rho_e$ than with the inertial length $d_e$. This result contrasts with the result in~\cite{alexandrova12} which showed no correlation ($C\sim 0.03$) between the spectral breaks (given by the used double-power-law model) and $\rho_e$. A possible explanation of the lack of the correlation in that work is the downsampling frequency of the STAFF-SA that does not allow one to capture properly the spectral breaks. We note that the moderate correlation with $d_e$ (0.58) maybe due to all intervals that have $\beta_e\sim 1$ (for those events $\rho_e\sim d_e$). We note finally that a similar correlation coefficient between $f_b$ and $\rho_e$ is found in magnetosheath turbulence ~\citep{huang13}, but a weaker correlation is found with $d_e$ ($C=0.27$). All these observations suggest the relevance of $\rho_e$ as a dissipation scale for the SW and in the magnetosheath.

% --------------------------------------------------------------------------------------------------------------------------------------------------------------------------
%%%%%%%%%%%%%%%%%			Nouvelle section             %%%%%%%%%%%%%%%%%%
%---------------------------------------------------------------------------------------------------------------------------------------------------------------------------
\section{Discussion}\label{discussion}
The results shown in the previous section underline the need to carefully handle the Cluster/STAFF data in the SW in order to analyze properly the energy spectra at electron scales. We showed in particular that in order to evidence the presence of the spectral breaks at electron scales time series with high SNR are needed. For the selected data a large number of magnetic energy spectra was found to have a clear break near the electron scale, followed by steep power-law-like spectra $f^{\alpha}$. The distribution of their slopes $\alpha$ was shown to range from $-3.5$ to $-5.5$ with a peak near $-4$. This distribution is wider than that of the slopes in the dispersive range which covers the values $-3.2 \lesssim \alpha \lesssim -2.3$. A comparison of these distributions to those reported in magnetosheath turbulence \citep{huang13} shows some interesting similarities and differences. In the dispersive range the distribution of the slopes is very similar: both are narrow and show a peak near $-2.8$, in general agreement with most of theoretical and numerical predictions on SW turbulence. However, in the electron dissipation range the magnetosheath spectra were found to be generally steeper with slopes $\alpha \sim -7$, whose distribution peaks near $-5.5$. This comparison stimulated a further analysis of the data to explain the difference between the SW and magnetosheath observations based on the differences in the SNRs between the two regions. Let us examine this effect in detail.

\begin{figure}
\includegraphics[height=6cm,width=7.5cm]{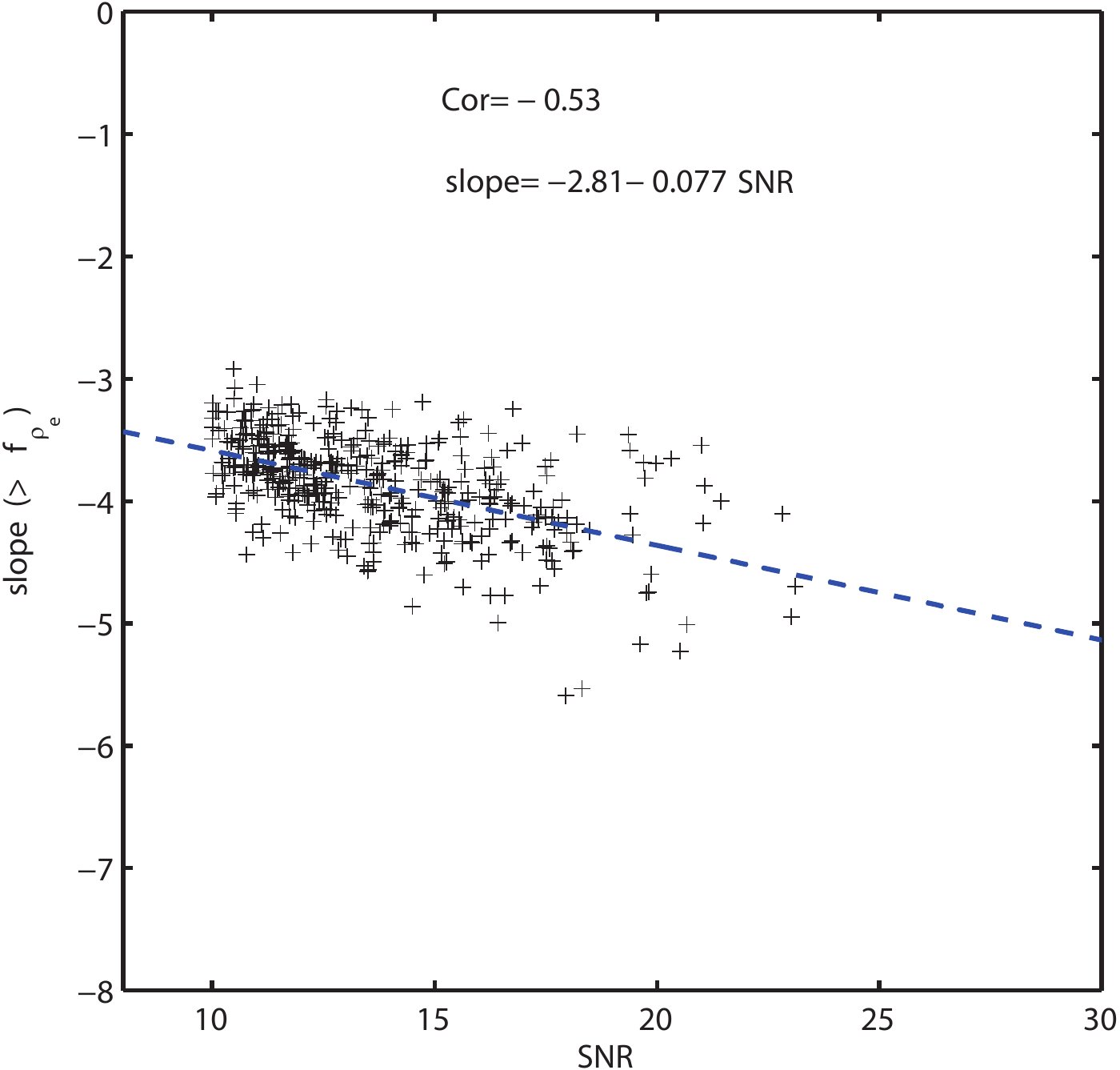}
\caption{Correlation between the frequencies of the observed spectral breaks $f_b$ with the SNR (see text) at $30$Hz of the studied magnetic energy spectra.\label{slopes_snr}}
\end{figure}

Although the data used here have the highest found SNR in ten years of data survey in the SW, it might be possible that even SNR values as high as $10$ are not sufficient enough to determine accurately the scaling above $f_b$. The reason is the limited extension of the spectrum above $f_b$ (typically less than one decade) due to the proximity of the sensitivity floor of the SCM. To test this hypothesis we plotted in Fig.~\ref{slopes_snr} the measured slopes of the spectra above $f_{\rho_e}$ as a function of the corresponding SNR. The figure shows a moderate correlation ($C=0.53$) between the SNR and the slopes. Typically most of the slopes $|\alpha|\lesssim3.5$, which were not observed in magnetosheath turbulence, correspond to SNR$<15$ and the highest values of the slopes $|\alpha|\gtrsim4.5$ correspond to SNR$\gtrsim15$. This correlation suggests that data with even higher SNR are needed to fully address the actual scaling of the magnetic energy spectra at sub-electron scales in the SW. Based on this conclusion the power-law fits and the interpretation of the low values of slopes ($|\alpha| \lesssim 3.5$) reported in this study should be considered with some caution. We emphasize that a weak correlation ($C<0.1$) is found in magnetosheath turbulence when SNR$>25$ ~\citep{huang13}.

\begin{figure}
\includegraphics[height=4cm,width=8.5cm]{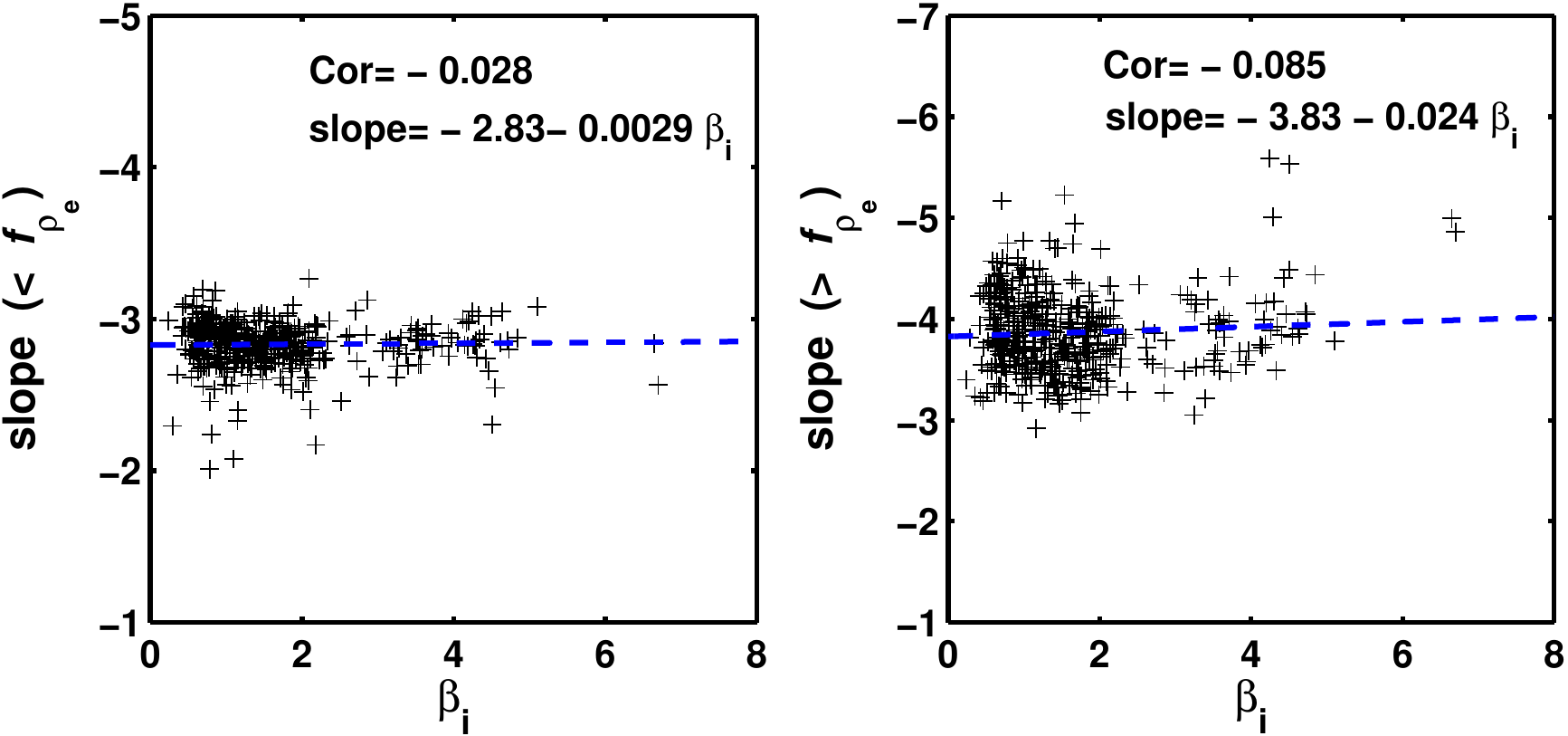}
\caption{Correlation between the frequencies of the observed spectral breaks $f_b$ with the plasma $\beta_i$ of all analyzed time intervals.\label{slopes_beta}}
\end{figure}

\begin{figure}
\includegraphics[height=4.5cm,width=7.5cm]{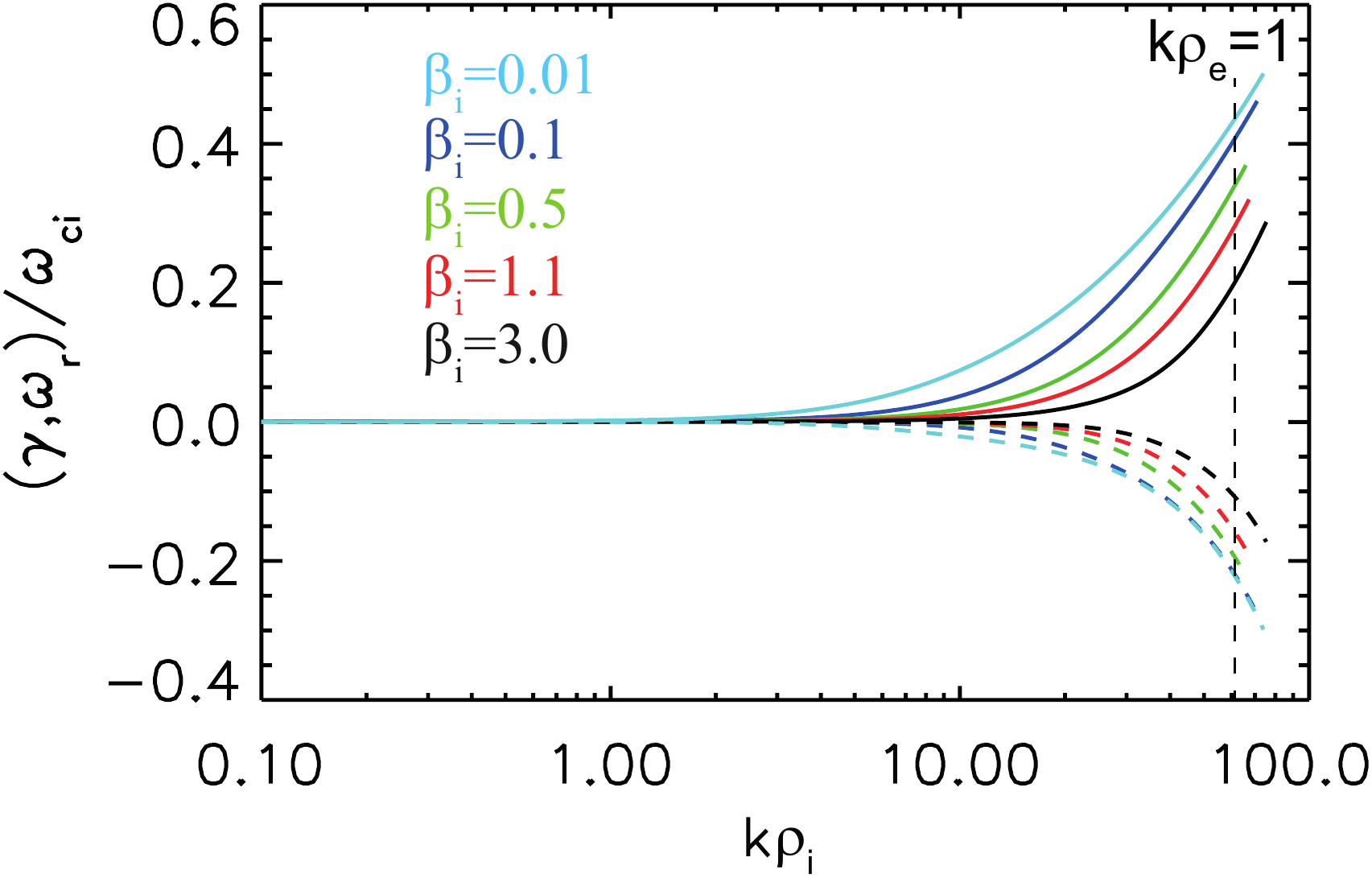}
\includegraphics[height=4.5cm,width=7.5cm]{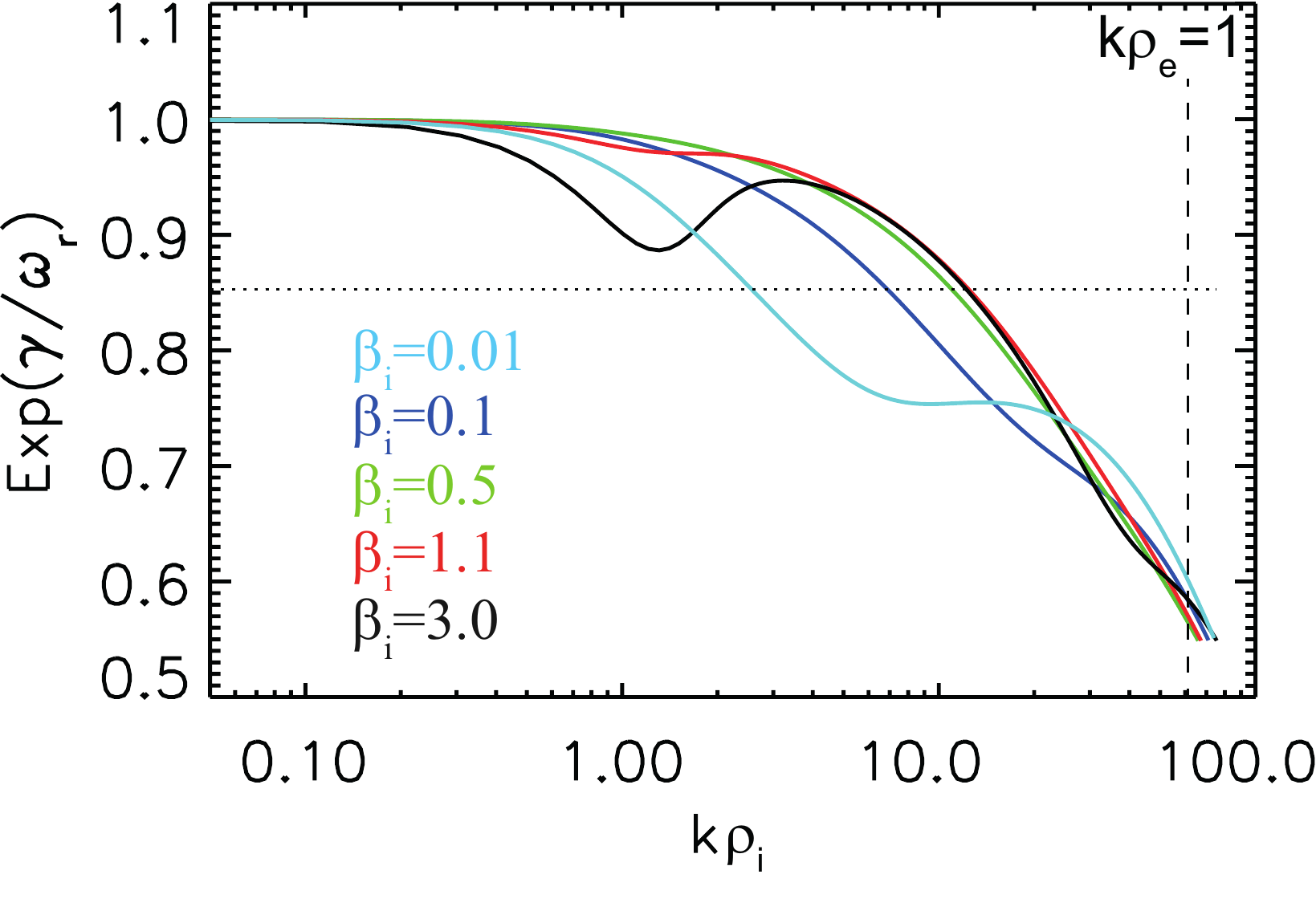}
\caption{Top: KAWs dispersion relations and dampings rates normalized to the proton gyrofrequency for $\theta_{\bf kB}=89.99^\circ$ and the given values of $\beta_i$. Bottom: the damping rates of the same modes normalized to one wave period. \label{disp}}
\end{figure}

Another interesting point that is worth a discussion is the possible dependence of the scaling of the magnetic energy spectra in the dispersive and the dissipation ranges on the plasma $\beta_i$. If one assumes that turbulence below the ion scale is dominated essentially by high oblique KAWs as suggested  by several SW observations \citep{bale05,sahraoui09, sahraoui10a,salem12,kiyani13} such a dependence could exist. This is indeed based on the known result from the linear kinetic theory that the damping of high oblique KAWs depends strongly on the plasma $\beta_i$. This can be seen clearly in Fig. \ref{disp} (top panel) which shows that the lower is the $\beta_i$ the more damped is the KAW mode (for the same angle of propagation). Based on this remark one would expect to observe a rapid decay of the energy spectra at low $\beta_i$, which would translate into steeper power-law or exponentially decaying spectra~\citep{howes11c}. To test this hypothesis we plotted on Fig.~\ref{slopes_beta} the correlation between the observed slopes below and above $f_{\rho_e}$ as a function of $\beta_i$. The results do not show any clear dependence between the values of the slopes and $\beta_i$. A possible explanation of that is the fact that the difference in the kinetic damping of the KAW modes is more pronounced between $\beta_i\sim 1$ and much lower values $\beta_i \sim 0.01$. This cannot be seen directly on Fig.~\ref{disp} (top). To evidence this fact plotted in Fig.~\ref{disp} (bottom) the damping of each KAW mode over one wave period. The figure shows that the modes with $\beta_i \gtrsim 0.5$ have nearly the same normalized damping rate at all scales, which is very different from the dampings of the modes at $\beta_i=0.01$ and $\beta_i=0.1$. These latters are shown to undergo a strong damping (defined when $2\pi \gamma/\omega_r \sim 1$) at respectively $k\rho_i\sim 3$ and $k\rho_i\sim 10$. This may explain the absence of correlation between the slopes and $\beta_i$ reported in Fig.~\ref{slopes_beta} since most of the observations reported here have $\beta_i \gtrsim 0.5$. Larger samples of SW data would allow one to obtain lower values of $\beta_i$ and to test the scenario proposed here. Other explanations are also possible such as the presence of non-local effects (due to large scales shear flows) that would overcome the strong damping at low $\beta_i$ and sustain the cascade down to the electron scales as suggested in~\cite{howes11c}. Another explanation is that KAW scenario discussed here may be irrelevant to interpret the present observations. In any case, the origin of the large spread of the slopes above $f_{\rho_e}$ reported here and in magnetosheath turbulence \citep{huang13} (as compared to those in the dispersive range) remains unclear and suggests the lack of universality of the turbulence at these small scales. The scaling of the spectra at sub-electron scales may depend on local electron plasma instabilities. Such dependence was indeed found between the slopes of the spectra at sub-ion scales and the local ion plasma instabilities~\citep{bale09}. Proving the existence of the same dependance at sub-electron scale requires having high time resolution measurements (a few ms) of electron distribution functions. Such measurements are not available with the current space missions. The NASA/MMS mission is expected to meet part of these requirements.

%---------------------------------------------------------------------------------------------------------------------------------------------------------------------------
%%%%%%%%%%%%%%%%%			Nouvelle section             %%%%%%%%%%%%%%%%%%
%---------------------------------------------------------------------------------------------------------------------------------------------------------------------------
\section{Conclusions}\label{conclusions}
We have surveyed more than ten years of the Cluster burst mode SCM data in the SW and focused on the magnetic energy spectra below the ion gyroscale $\rho_i$. We discussed several instrumental caveats that potentially may influence the estimation of the scaling of the spectra at electron scales. We showed in particular that high SNRs are needed in order to determine properly the scaling of the energy spectra. Once all the caveats were considered we found a large number of spectra having clear spectral breaks near the Taylor-shifted electron gyroscale $f_{\rho_e}$. A double-power-law model has been adopted to fit the spectra. The distribution of the slopes below the spectral break  $f_b$ was found to be narrow and centered around $\sim -2.8$, while that above the $f_b$ was found wider and centered near the value $\sim -4$. The peak of the distribution of the slopes below $f_b$ is consistent with similar observations of SW and magnetosheath turbulence \citep{alexandrova12,huang13} and with most of the existing theoretical or numerical predictions. The spectra above $f_b$ are however shallower than those reported in magnetosheath turbulence. A possible explanation of that discrepancy is the difference in the SNR between the SW and the magnetosheath. A moderate correlation was indeed  found between small values of the slopes, typically $|\alpha|<4$, with lower values of SNR. The reported wide distribution of the slopes combined with earlier observations that showed exponential-like scaling indicate that the physics of the electron scales is possibly not universal and might be controlled by several plasma parameters. Based on the assumption that KAW turbulence dominates at electron scales and on the fact that the linear damping of the KAW modes (from the Vlasov-Maxwell equations) strongly depends on $\beta_i$, an attempt to find a correlation between the observed slopes and the plasma $\beta_i$ was carried-out, but the result was not conclusive. 

The Cluster-Search Coil data allow for unprecedented in-depth studies of electron scales turbulence in the SW. However, as shown in this study, magnetic field data with higher SNR is now needed to fully and unambiguously characterize the turbulence cascade and dissipation at electron scales in the SW. Besides that one needs furthermore to have accurate measurements of electric field fluctuations and high time resolution of ions and electrons distribution functions, which are not available on current space missions. A new mission, TOR, is currently being designed to fulfill these requirements. TOR should allow in the upcoming years to fully resolve the problem of turbulence cascade and dissipation at electron scales in the SW. 
\acknowledgments
This work is part of the project THESOW funded by L'Agence Nationale de la Recherche (ANR, France). J. De Patoul is funded through the ANR grant. The FGM and CIS data come from the CAA (ESA) and AMDA (CDPP, IRAP, France). FS thanks O. Gurcan for fruitful discussions.
\email{fouad.sahraoui@lpp.polytechnique.fr}

%\section{Appendix material}




\clearpage

\clearpage


\begin{thebibliography}{}

\bibitem[Alexandrova et al.(2009)]{alexandrova09} Alexandrova, O., Saur, J., Lacombe, C., Mangeney, A., Mitchell, J., Schwartz, S. J., \& Robert, P. 2009, \prl, 103, 165003
%\bibitem[Alexandrova et al.(2010)]{alexandrova10} Alexandrova, O., Saur, J., Lacombe, C., Mangeney, A., Schwartz, S. J., Mitchell, J., Grappin, R.,  \& Robert, P. 2010, Solar Wind 12, AIP Conference proceedings, 1612.
\bibitem[Alexandrova et al.(2012)]{alexandrova12} Alexandrova, Lacombe, C., Mangeney, A., Grappin, R., \& Maksimovic, M. 2012, \apj, 760, 121
\bibitem[Bale et al.(2005)]{bale05} Bale, S. D. , Kellogg, P. J. , Mozer, F. S. , Horbury, T. S. , \& R\`eme, H. 2005, \prl, 94, 215002
\bibitem[Bale et al.(2009)]{bale09} Bale, S. D. , Kasper, J. C., Howes, G. G., Quataert, E., Salem, C., \& Sundkvist, D. 2009, \prl, 103, 211101
\bibitem[Biskamp et al.(1999)]{biskamp99} Biskamp, D., Schwarz, E., Zeiler, A, Celani, A., \& Drake, J. F.  1999, Phys. Plasmas, 6, 751
\bibitem[Boldyrev \& Perez (2012)]{boldyrev12} Boldyrev, S., \& Perez, J. C., 2012, \apj, 758, L44
\bibitem[Bourouaine et al.(2012)]{bourouaine12} Bourouaine, S., Alexandrova, O., Marsch, E., \& Maksimovic, M. 2012, \apj , 749
\bibitem[Camporeale \& Burgess(2011)]{camporeale11} Camporeale, E., \& Burgess., D. 2011, \apj, 730, 114
\bibitem[Chang et al.(2011)]{chang11} Chang, O., Gary, S. P., Wang, J 2011, \grl, 38, L22102, doi:10.1029/2011GL049827
\bibitem[Chen et al.(2011)]{chen10} Chen, C. H. K. , Horbury,T. S., Schekochihin, A. A., Wicks, R. T., Alexandrova, O., \& Mitchell, J. 2010, \prl, 104, 255002
\bibitem[Cornilleau et al.(2003)]{cornilleau03} Cornilleau-Wehrlin, N. et al. 2003, Ann. Geophys. 21, 437-456
%\bibitem[Dmitruk et al.(2009)]{dmitruk09} Dmitruk, P., \& Matthaeus, W. H. 2009, Phys. Plasmas, 16, 062304
%\bibitem[Galtier (2006)]{galtier06} Galtier, S. 2006, J. Plasma Phys.,  72, 721
%\bibitem[Galtier (2008)]{galtier08} Galtier, S. 2008, \pre  77, 015302
\bibitem[Gary \& Smith(2009)]{gary09} Gary, S. P. \& Smith, C. W. 2009, \jgr, 114, A12105
%\bibitem[Goldreich \& Sridhar(1995)]{goldreich95} Goldreich, P. \& Sridhar, S. 1995, \apj, 438, 763
%\bibitem[Goldstein et al.(1994)]{goldstein94} Goldstein, M. L., Roberts, M. D.,  \& Fitch, C. 1994, \jgr, 99, 1005
\bibitem[Gurcan et al.(2009)]{gurcan09} Gurcan,O. D., X. Garbet, P. Hennequin, P. H. Diamond, A. Casati,\& G. L. Falchetto, \prl, 102, 255002.
\bibitem[Hasegawa et al.(1978)]{hasegawa78} Hasegawa, A., T. Imamura, K. Mima,\& T. Taniuti, JPSJ, 43, 255002.
%\bibitem[Hellinger et al.(2006)]{hellinger06} Hellinger, P., Tr\'avn\'icek, P., \& Lazarus, A. J. 2006, \grl, 33, L09101
\bibitem[Kiyani et al.(2009)]{kiyani09} Kiyani, K. H.,  Chapman, C., Khotyaintsev, Yu.V., Dunlop, M.W., \& Sahraoui, F. 2009, \prl, 103, 075006
\bibitem[Kiyani et al.(2013)]{kiyani13} Kiyani, K. H.,  Chapman, C., Sahraoui, F., Hnat, B., Fauvarque, O., \& Khotyaintsev, Yu.V.,  2013, \apj, 763,10
%\bibitem[Krauss-varban et al.(1994)]{krauss94} Krauss-Varban, D., Omidi, N., \& Quest, K., B. 1994, \jgr, 99, 5987
%\bibitem[Lacombe et al.(1995)]{lacombe95} Lacombe, C., Belmont, G., Hubert, D., Harvey, C. C., Mangeney, A., Russell, C. T., Gosling, J. T., \& Fuselier, S. A. 1995, Ann. Geophys. 13, 343
%\bibitem[Leamon et al.(1998)]{leamon98} Leamon, R. J.,  Smith, C. W., Ness, N. F. , Matthaeus, W. H. \& Wong, H. K. 1998, \jgr, 103, 4775
\bibitem[Leamon et al.(1999)]{leamon99} Leamon, R. J.,  Smith, C. W., \& Ness, N. F.  1999, \jgr, 104, 22331
%\bibitem[Li \& Habbal(2001)]{li01} Li, X., \& Habbal, S. R. 2001, \jgr, 106, 10669
\bibitem[Meyrand \& Galtier(2010)]{meyrand10} Meyrand, R., \& Galtier, S. 2010, \apj, 721, 1421-1424
\bibitem[Meyrand \& Galtier(2012)]{meyrand12} Meyrand, R., \& Galtier, S. 2012, \prl, 109, 194501
\bibitem[Mazelle (2012)]{mazelle12} Mazelle, C.,  Private communication
%\bibitem[Howes et al.(2008a)]{howes08a} Howes, G. 2008a, Phys.  Plasmas, 15, 055904
%\bibitem[Howes et al.(2008a)]{howes08a} Howes, G., 2008a, \prl, 101, 175005
%\bibitem[Howes et al.(2008b)]{howes08b} Howes, G.,  Cowley, S. C.,  Dorland, Hammett, G. W., Quataert, E., \& Schekochihin, A., 2008b, \jgr, 113, A05103
\bibitem[Howes (2009)]{howes09} Howes, G., 2009, Nonlin. Processes Geophys., 16, 219
\bibitem[Howes et al.(2011a)]{howes11a} Howes, G., TenBarge, J. M., Dorland, W., Quataert, E., Schekochihin, A., Numata, R., \& Tatsuno, T. 2011, \prl, 107, 035004
%\bibitem[Howes et al.(2011b)]{howes11b} Howes, G.,  Bale, S. D. , Klein, K. G. , Chen,  C. H. K., Salem, C., \& TenBarge 2011, arXiv:1106.4327v1 [astro-ph.SR]
\bibitem[Howes et al.(2011c)]{howes11c} Howes, G., TenBarge, J. M., Dorland, W. 2011, Phys. Plasmas, 18, 102305
\bibitem[Huang et al.(2013)]{huang13} Huang, S., Sahraoui, S. 2012, in preparation
%\bibitem[Narita et al.(2010a)]{narita10a} Narita, Y., Glassmeier, K. H., Sahraoui, F., \& Goldstein, M. L. 2010a, \prl, 104, 171101
\bibitem[Matthaeus et al.(2008)]{matthaeus08} Matthaeus, W. H., Servidio, S., \& Dmitruk, P. 2008, \prl, 101, 149501
%\bibitem[Narita et al.(2010b)]{narita10b} Narita, Y., Sahraoui, F., Goldstein, M. L., \& Glassmeier, K. H. 2010b, \jgr, 115, A04101, doi:10.1029/2009JA014742
%\bibitem[Narita et al.(2011)]{narita11} Narita, Y., Gary, S. P., Saito, S., Glassmeier, K. H., \& Motschmann, U 2011, \grl, 38, L05101
\bibitem[Podesta et al.(2010)]{podesta10} Podesta, J. J., Borovsky, J. E., Gary, \& S.P. 2010, \apj, 712, 685
%\bibitem[Pin\c{c}on \& Lefeuvre(1991)]{pincon91} Pin\c{c}on, J.L., \& Lefeuvre L. 1991, \jgr, 96, 1789
%\bibitem[Sahraoui et al.(2003a)]{sahraoui03a} Sahraoui, F. et al., 2003a, \jgr,108, 1335
\bibitem[Sahraoui et al.(2003b)]{sahraoui03b} Sahraoui, F., Belmont, G., Rezeau, L. 2003b, Phys. Plasmas, 10, 1325
\bibitem[Sahraoui et al.(2004)]{sahraoui04} Sahraoui, F. et al., 2004, Ann. Geophys. 22, 2283
\bibitem[Sahraoui et al.(2006)]{sahraoui06} Sahraoui, F., Belmont,G., Rezeau, L., Cornilleau-Wehrlin, N., Pin\c{c}on, J. L. , \& Balogh, A. 2006, \prl, 96, 075002
%\bibitem[Sahraoui et al.(2007)]{sahraoui07} Sahraoui, F., Galtier, S., Belmont,G. 2007, J. Plasma Phys., 73, 723
\bibitem[Sahraoui et al.(2009)]{sahraoui09} Sahraoui, F., Goldstein, M. L. , Robert, P. , Khotyaintsev, Y. V.  2009, \prl, 102, 231102
\bibitem[Sahraoui et al.(2010a)]{sahraoui10a} Sahraoui, F.,  Goldstein, M. L., Belmont, G., Canu, P., \& Rezeau, L. 2010a, \prl, 105, 131101
%\bibitem[Sahraoui et al.(2010b)]{sahraoui10b} Sahraoui, F.,  Belmont, G., Goldstein, M. L., \& Rezeau, L. 2010b, \jgr, 115, A04206
\bibitem[Sahraoui et al.(2010c)]{sahraoui10c} Sahraoui, F.,  Goldstein, M. L., Belmont, G., Roux, A., Rezeau, L., Canu, P., Robert, P., Cornilleau-Wehrlin, N., Le Contel, O., Dudok De Wit, T., Pin\c{c}on, J.L., \& Kiyani, K. 2010c, Plant. Space Science, 59, 585
%\bibitem[Sahraoui et al.(2011)]{sahraoui11} Sahraoui, F.,  Huang, S.,  \& Goldstein, M. L., In preparation
\bibitem[Sahraoui et al.(2012)]{sahraoui12} Sahraoui, F.,  Belmont, G.,  \& Goldstein, M. L., \apj, 748, 100, doi:10.1088/0004-637X/748/2/100
\bibitem[Salem et al.(2012)]{salem12} Salem,C. S. , G. G. Howes, D. Sundkvist, S. D. Bale, C. C. Chaston, C. H. K. Chen, and F. S. Mozer, \apjl, 745, L9, doi:10.1088/2041-8205/745/1/L9
\bibitem[Schekochihin et al.(2009)]{schekochihin09} Schekochihin, A., Cowley, S. C., Dorland, W., Hammett, G. W., Howes, G. G., Quataert, E., \& Tatsuno, T. 2009, \apjs, 182, 310
%\bibitem[Shebalin et al.(1983)]{shebalin83} Shebalin, J.V., Matthaeus, W. H.,  \& Montgomery, D. 1983, J. Plasma Phys., 29
%\bibitem[Song et al.(1994)]{song94} Song, P, Russell, C.T.,  \& Gary, S. P. 1994, \jgr, 99, 6011
%\bibitem[Stawicki et al.(2001)]{stawicki01} Stawicki, O., Gary, S. P., \& Li, H. 2001, \jgr, 106, 8273
%\bibitem[Tjulin et al.(2005)]{tjulin05} Tjulin, A., Pin\c{c}on, J.L., Sahraoui, F., Andr\'e, M., \& Cornilleau-Wehrlin, N. 2005, \jgr, 110, A11224, doi:10.1029/2005JA011125
%\bibitem[R\"{o}nnmark(1982)]{ronmark82} R\"{o}nnmark, K. 1982, Kiruna Geophysical Institute Report, 179

\end{thebibliography}
\end{document}